\documentclass[twocolumn]{jpsj2} 
\usepackage{bm}
\usepackage{amsfonts}
\def\eq{\hspace{-3mm}&=&\hspace{-3mm}}
\def\espace{\hspace{-3mm}&&\hspace{-3mm}}
\def\eequiv{\hspace{-3mm}&\equiv&\hspace{-3mm}}

\def\ket{\rangle}
\def\d{\mbox{\rm d}}
\def\e{\mbox{\rm e}}
\def\i{\mbox{\rm i}}
\def\smalli{\mbox{\scriptsize \rm i}}

\title{Matter-Wave Solitons in an $F=1$ Spinor Bose--Einstein Condensate}

\author{Jun'ichi \textsc{Ieda}$^{1}$\thanks{E-mail address: ieda@monet.phys.s.u-tokyo.ac.jp},
Takahiko \textsc{Miyakawa}$^{1,2}$\thanks{E-mail address: tmiyakawa@optics.arizona.edu} and
Miki \textsc{Wadati}$^{1}$}

\inst{$^{1}$Department of Physics, Graduate School of Science,
University of Tokyo, Bunkyo-ku, Tokyo 113-0033 \\
$^{2}$Optical Sciences Center, University of Arizona, Tucson, AZ 85721, USA}

\abst{
Following our previous work
[J. Ieda, T. Miyakawa, M. Wadati, cond-mat/0404569]
on a novel integrable model describing soliton dynamics of
an $F=1$ spinor Bose--Einstein condensate,
we discuss in detail the properties of the multi-component system
with spin-exchange interactions.
The exact multiple bright soliton solutions are obtained for
the system where the mean-field interaction is attractive ($c_0 < 0$) 
and the spin-exchange interaction is ferromagnetic ($c_2 < 0$).
A complete classification of the one-soliton solution with respect
to the spin states
and an explicit formula of the two-soliton solution are presented.
For solitons in polar state, there exists a variety of different shaped
solutions including twin peaks. We show that a ``singlet pair" density
can be used to distinguish those energetically degenerate solitons.
We also analyze collisional effects between solitons in the same or
different spin state(s) by computing the asymptotic forms of their initial
and final states. The result reveals that it is possible to manipulate
the spin dynamics by controlling the parameters of colliding solitons.
}

\kword{Bose--Einstein condensate, spin degrees of freedom, bright soliton,
nonlinear Schr\"odinger equation, inverse scattering method, atom optics}

\begin{document}
\maketitle

\section{\label{sec:introduction}Introduction}
Solitons are a universal feature of nonlinear phenomena~\cite{refsoliton}
in many different physical areas such as particle physics, nonlinear optics,
plasma physics, fluid dynamics, and condensed matter physics.
Recently, two teams in the US~\cite{SolRice} and France~\cite{SolEns}
performed experiments that demonstrated matter-wave bright solitons with
Bose--Einstein condensates (BECs) of gaseous ${}^7{\rm Li}$ atoms.
In general, solitons are formed under the balance between nonlinearity and
dispersion.
For atomic BECs whose macroscopic wave functions obey the Gross--Pitaevskii (GP)
equation or nonlinear Schr\"odinger (NLS) equation, the former is attributed
to the interatomic interactions while the latter comes from the kinetic energy.

Either dark or bright solitons are allowable depending on the positive or negative
sign of the interatomic coupling constants, respectively.
Bright solitons created in the experiments are themselves condensates and
propagated over much larger distances than dark solitons which, on the other hand,
can only exist as notches or holes within the condensate itself~\cite{darksolex}.
Those observed matter-waves behaved like scalar fields
since the spins of condensed atoms are frozen under additional magnetic fields.
In line with this, many theoretical studies on the bright soliton
formation and propagation of attractive BECs have been done mainly in the single
component systems~\cite{Khawaja,Salasnich,Leung,Carr}.

Multi-component generalization of the soliton dynamics is very natural
in the context of atomic BECs because of several ways to create
such systems, i.e., the mixtures with two different atomic species/hyperfine
states~\cite{binaryBEC} and the internal degrees of freedom
liberated under an optical trap~\cite{spinor_BEC_MIT}.
Moreover, when one explores future applications of matter-wave solitons
in atom optics~\cite{Meystre},
such as atom laser, atom interferometry, and coherent atom transport,
the multiple condensates have various potential utilities, e. g., the multi-channel
signals and their switching.

So far, multiple solitons were theoretically investigated
in two-component condensates~\cite{2component}
and also in an arbitrary number components system~\cite{Soljacic};
the nonlinearity in both cases consists only of the intensities, i.e.,
the squared absolute values of the components.
The coupled NLS equations with such nonlinearity, often referred
to as the intensity-coupled type, are extensively used for
modeling multi-mode optical solitons~\cite{opt_soliton_review}.
In contrast, the so-called spinor condensates~\cite{Ohmi,Ho,Law,Koashi,Ciobanu,Ueda},
being another candidate that supports multiple matter-wave solitons,
have nontrivial nonlinear terms reflecting the $SU(2)$ symmetry of the spins.
The spin-exchange interactions that are the sources of the spin-mixing
within condensates~\cite{spindynamics_th,spindynamics_ex} are the exceptions of
the intensity-coupled nonlinearity and there is no analogue in conventional optics.

In our previous work~\cite{IMWlett}, we discovered a novel integrable model
which describes the dynamics of a spinor BEC 
and derived the multiple $N$-soliton solution.
We assume specifically an atomic condensate in the $F=1$ hyperfine spin state
confined in one-dimensional space.
In all-optical dipole traps, the three spin substates
$|F=1,m_F=1\ket$, $|F=1,m_F=0\ket$, and $|F=1,m_F=-1\ket$,
where $m_F$ is the magnetic quantum number, are essentially free and
the spinor nature of the alkali atoms can be manifested.
In this paper, we discuss the soliton properties of this system
in the framework of the integrable model, which brings us to understand the
intrinsic aspects of this complex nonlinear system and provides a rigid
reference for further numerical and experimental investigations.

The paper is organized as follows. In \S~\ref{sec:model}, the model of
spinor condensate is presented, and its embedding into the integrable
equation is reviewed.
In \S~\ref{sec:ISM}, we briefly summarize the inverse scattering method,
the way to the exact $N$-soliton solutions of this model, and 
provide some conserved quantities.
Next, in \S~\ref{sec:OSS}, the properties of the one-soliton solution
classified  by the spin states are investigated in detail. In particular,
we introduce a singlet pair density to resolve energetically degenerate
solitons in the polar state.
Then, in \S~\ref{sec:switch}, collisional effects between solitons 
in the same or different spin state(s) are examined by analyzing
the asymptotic forms of their initial and final state.
As one of the most significant phenomena of the two-soliton collisions,
spin switching is predicted.
Section~\ref{sec:conclusions} concludes the paper.
In Appendix, we give the explicit formula of general two-soliton solutions
in this spinor model.

\section{\label{sec:model}$F=1$ Spinor Condensate}
Atoms in the $F=1$ state are characterized by a vectorial field
operator with the components subject to the hyperfine spin manifold.
The three-component field $\hat{{\bm \Psi}}=\{\hat{\Psi}_{1},
\hat{\Psi}_{0},\hat{\Psi}_{-1}\}^T$,
with the superscript $T$ denotes the transpose,
satisfies the bosonic commutation relations:
\begin{equation}
[\hat{\Psi}_\alpha(x,t),\hat{\Psi}^{\dagger}_\beta(x^\prime,t)]
=\delta_{\alpha,\beta}\delta(x-x^\prime).
\end{equation}
Here we assume that the system is one dimensional (1D): 
the trap is suitably anisotropic such that the transverse
spatial degrees of freedom is factorized from the longitudinal
and all the hyperfine states are in transverse ground state.
Those quasi-one dimensional regime can be achieved experimentally
(see the discussion in refs. 8).
The interaction between atoms in the $F=1$ hyperfine state
is given by~\cite{Ho,Ohmi} 
\begin{equation}
\label{int}
\hat{V}(x_1-x_2)=\delta(x_1-x_2)(c_0+c_2\hat{{\bm F}}_1\cdot\hat{{\bm F}}_2),
\end{equation}
where $\hat{{\bm F}}_i$ are the spin operators of two atoms.
In this expression,
\begin{eqnarray}
c_0\eq4\pi\hbar^2(a_0+2a_2)/3m, \nonumber\\
c_2\eq4\pi\hbar^2(a_2- a_0)/3m, 
\end{eqnarray}
where $a_f$ are the s-wave scattering lengths for the channel of
total hyperfine spin $f$ and $m$ is the mass of the atom.
As a result of the symmetry required
for bosonic atoms, it can be shown that only even-$f$ states
contribute to $\hat{V}(x_1-x_2)$. Therefore, in the case of $F=1$ atoms,
there are two scattering channels: $f=2$ for the parallel spin collision
and $f=0$ for the antiparallel.
According to ref. 24,
the effective 1D couplings $\bar{c}_{0,2}$ are represented by
\begin{equation}
\bar{c}_0=c_0/2a_\perp^2,\quad \bar{c}_2=c_2/2a_\perp^2,
\end{equation}
where $a_\perp$ is the size of the  transverse ground state. 

Thus, the second-quantized Hamiltonian is
\begin{eqnarray}
H\eq \int \d x \,\bigg(\frac{\hbar^2}{2m}\partial_x\hat{\Psi}^\dagger_\alpha
\cdot\partial_x\hat{\Psi}_\alpha+
\frac{\bar{c}_0}{2}\hat{\Psi}^\dagger_\alpha
\hat{\Psi}^\dagger_{\alpha^\prime} \hat{\Psi}_{\alpha^\prime}
\hat{\Psi}_\alpha \nonumber \\
\espace \quad\quad\quad{}+\frac{\bar{c}_2}{2}\hat{\Psi}^\dagger_\alpha
\hat{\Psi}^\dagger_{\alpha^\prime}{\bm f}^T_{\alpha \beta} \cdot
{\bm f}_{\alpha^\prime \beta^\prime}\hat{\Psi}_{\beta^\prime}
\hat{\Psi}_{\beta}\bigg),
\end{eqnarray}
where repeated subscripts $\{\alpha,\beta,\alpha',\beta'=1,0,-1\}$
should be summed up.
The explicit form of the above Hamiltonian is shown to be
\begin{eqnarray}
H\eq \int \d x\, \bigg\{\frac{\hbar^2}{2m}\partial_x\hat{\Psi}^\dagger_\alpha
\cdot\partial_x\hat{\Psi}_\alpha+\frac{\bar{c}_0+\bar{c}_2}{2}
\Big[\hat{\Psi}^\dagger_1 \hat{\Psi}^\dagger_1
\hat{\Psi}_1 \hat{\Psi}_1 \nonumber\\
\espace {}+\hat{\Psi}^\dagger_{-1}
\hat{\Psi}^\dagger_{-1} \hat{\Psi}_{-1} \hat{\Psi}_{-1}
+2\hat{\Psi}^\dagger_0 \hat{\Psi}_0(\hat{\Psi}^\dagger_{1}
\hat{\Psi}_1+\hat{\Psi}^\dagger_{-1}\hat{\Psi}_{-1})\Big]
\nonumber\\
\espace {}+(\bar{c}_0-\bar{c}_2) \hat{\Psi}^\dagger_1 \hat{\Psi}^\dagger_{-1}
\hat{\Psi}_{-1} \hat{\Psi}_1
+\frac{\bar{c}_0}{2}\hat{\Psi}^\dagger_0 \hat{\Psi}^\dagger_0
\hat{\Psi}_0 \hat{\Psi}_0
\nonumber\\
\espace {}+\bar{c}_2 \left[\hat{\Psi}^\dagger_1 \hat{\Psi}^\dagger_{-1}
\hat{\Psi}_0 \hat{\Psi}_0+\hat{\Psi}^\dagger_0 \hat{\Psi}^\dagger_0
\hat{\Psi}_{-1} \hat{\Psi}_1\right]\bigg\},
\end{eqnarray}
where 
the following expression of spin-1 matrices
${\bm f}= \{f^x,\,f^y,\,f^z \}^T$
is employed,
\begin{eqnarray}
\label{spin1 mtx}
f^x\eq\frac{1}{\sqrt{2}}\left(\hspace*{-1mm}\begin{array}{ccc}
0\hspace*{-1mm}&\hspace*{-1mm}1\hspace*{-1mm}&\hspace*{-1mm}0 \\
1\hspace*{-1mm}&\hspace*{-1mm}0\hspace*{-1mm}&\hspace*{-1mm}1 \\
0\hspace*{-1mm}&\hspace*{-1mm}1\hspace*{-1mm}&\hspace*{-1mm}0
\end{array}\hspace*{-1mm}\right),
f^y=\frac{\i}{\sqrt{2}}\left(\hspace*{-1mm}\begin{array}{ccc}
0\hspace*{-1mm}&\hspace*{-1mm}-1\hspace*{-1mm}&\hspace*{-1mm} 0 \\
1\hspace*{-1mm}&\hspace*{-1mm} 0\hspace*{-1mm}&\hspace*{-1mm}-1 \\
0\hspace*{-1mm}&\hspace*{-1mm} 1\hspace*{-1mm}&\hspace*{-1mm} 0
\end{array}\hspace*{-1mm}\right),
\nonumber\\
f^z\eq\left(\hspace*{-1mm}\begin{array}{ccc}
1\hspace*{-1mm}&\hspace*{-1mm}0\hspace*{-1mm}&\hspace*{-1mm} 0 \\
0\hspace*{-1mm}&\hspace*{-1mm}0\hspace*{-1mm}&\hspace*{-1mm} 0 \\
0\hspace*{-1mm}&\hspace*{-1mm}0\hspace*{-1mm}&\hspace*{-1mm}-1
\end{array}\hspace*{-1mm}\right).
\end{eqnarray}

In the mean-field theory of BECs, the three-component condensate wave function
${\bm\Phi}(x,t)$ is obtained by
\begin{eqnarray}
{\bm\Phi}(x,t)\eequiv \langle \hat{\bm \Psi}(x,t) \rangle\nonumber\\
\eq\left\{\Phi_1(x,t),\Phi_0(x,t),\Phi_{-1}(x,t)\right\}^T,
\end{eqnarray}
which is normalized to the total number of atoms $N_{\rm T}$:
\begin{equation}
\int \d x \,{\bf \Phi}(x,t) ^{\dagger} \cdot {\bf \Phi}(x,t)
=N_{\rm T}.
\end{equation}

The time-evolution of spinor condensate wave function ${\bm \Phi}(x,t)$
can be derived from the variational principle:
\begin{equation}
\label{variation}
\i\hbar\partial_t \Phi_\alpha(x,t)=\frac{\delta E_{\rm GP}}
{\delta \Phi^*_\alpha(x,t)},
\end{equation}
where the Gross--Pitaevskii energy functional is given by
\begin{eqnarray}
\label{MFenergy}
E_{\rm GP}\eq\int \d x\, \bigg\{\frac{\hbar^2}{2m}\partial_x\Phi^*_\alpha
\cdot\partial_x\Phi_\alpha
+\frac{\bar{c}_0+\bar{c}_2}{2}\Big[|\Phi_1|^4+|\Phi_{-1}|^4\nonumber\\
\espace{}+2|\Phi_0|^2(|\Phi_1|^2+|\Phi_{-1}|^2)\Big]
+(\bar{c}_0-\bar{c}_2)|\Phi_1|^2|\Phi_{-1}|^2
\nonumber\\
\espace{}+\frac{\bar{c}_0}{2}|\Phi_0|^4+\bar{c}_2(\Phi^*_1\Phi^*_{-1}\Phi^2_0
+{\Phi^*_0}^2\Phi_1\Phi_{-1})\bigg\}.
\end{eqnarray}
Substituting eq.(\ref{MFenergy}) into eq.(\ref{variation}),
we get a set of equations for the spinor condensate
wave functions:
\begin{eqnarray}
\i\hbar\partial_t\Phi_1\eq -\frac{\hbar^2}{2m}\partial^2_x\Phi_1+
(\bar{c}_0+\bar{c}_2)\{|\Phi_1|^2+|\Phi_{0}|^2\}\Phi_1\nonumber\\
\espace{}+(\bar{c}_0-\bar{c}_2)|\Phi_{-1}|^2\Phi_1
+\bar{c}_2\Phi^*_{-1}\Phi^2_0,\nonumber\\
\i\hbar\partial_t\Phi_0\eq
-\frac{\hbar^2}{2m}\partial^2_x\Phi_0+(\bar{c}_0+\bar{c}_2)
\{|\Phi_1|^2+|\Phi_{-1}|^2\}\Phi_0\nonumber\\
\espace{}+\bar{c}_0|\Phi_0|^2\Phi_0
+2\bar{c}_2\Phi^*_0\Phi_1\Phi_{-1},\nonumber\\
\label{TDeq-1}
\i\hbar\partial_t\Phi_{-1}\eq-\frac{\hbar^2}{2m}\partial^2_x\Phi_{-1}
+(\bar{c}_0+\bar{c}_2)\{|\Phi_{-1}|^2+|\Phi_{0}|^2\}\Phi_{-1}\nonumber\\
\espace{}+(\bar{c}_0-\bar{c}_2)|\Phi_{1}|^2\Phi_{-1}
+\bar{c}_2\Phi^*_{1}\Phi^2_0.
\end{eqnarray}

To analyze the dynamical properties of these coupled system,
we employ the integrable model found in ref.~23.
We consider the system with the integrable condition
of coupling constants $\bar{c}_0=\bar{c}_2\equiv-c<0$, equivalently
scattering lengths $2a_0=-a_2>0$.
This situation corresponds to attractive mean-field interaction
and ferromagnetic spin-exchange interaction.
The effective interactions between atoms in a BEC can be
tuned with the so-called Feshbach resonance~\cite{Mag1,Mag2}.
Moreover, alternative techniques, which do not affect
the rotational symmetry of the internal spin states, 
such as optically induced
Feshbach resonance~\cite{Opt1,Opt2,Opt3} are available.

In the dimensionless form:
${\mathbf \Phi}\to \{\phi_1, \sqrt{2}\phi_0,\phi_{-1}\}^T$,
where time and length are measured respectively
in units of $\bar{t}=\hbar a_\perp/c$ and $\bar{x}=\hbar\sqrt{a_\perp/2mc}$,
we rewrite eqs.(\ref{TDeq-1}) as follows,
\begin{eqnarray}
\hspace*{-8mm}&&\i\partial_t \phi_1= -\partial^2_{x}\phi_1
-2\{|\phi_1|^2+2|\phi_0|^2\}\phi_1-2\phi^*_{-1}\phi^2_0,\nonumber\\
\hspace*{-8mm}&&\i\partial_t \phi_0= -\partial^2_{x}\phi_0
-2\{|\phi_{-1}|^2+|\phi_0|^2+|\phi_1|^2\}\phi_0-2\phi^*_0\phi_1\phi_{-1},
\nonumber\\
\label{stdeq-1}
\hspace*{-8mm}&&\i\partial_t \phi_{-1}= -\partial^2_{x}\phi_{-1}
-2\{|\phi_{-1}|^2+2|\phi_0|^2\}\phi_{-1}-2\phi^*_{1}\phi^2_0.
\end{eqnarray}
These coupled equations are equivalent to a $2 \times 2$ matrix version of
nonlinear Schr\"{o}dinger (NLS) equation:
\begin{equation}
\label{NLS}
\i\partial_t Q+ \partial^2_x Q+2QQ^\dagger Q={\it O},
\end{equation}
with an identification,
\begin{equation}
\label{reduction}
Q=\left(\hspace*{-1mm}\begin{array}{cc}
\phi_1 \hspace*{-1mm}&\hspace*{-1mm} \phi_0 \\
\phi_0 \hspace*{-1mm}&\hspace*{-1mm} \phi_{-1}
\end{array}\hspace*{-1mm}
\right).
\end{equation}
Since the matrix NLS equation (\ref{NLS}) is completely
integrable~\cite{Tsuchida1}, the integrability of the reduced
equations (\ref{stdeq-1}) are proved automatically~\cite{IMWlett}.
Remark that the general $M\times L$ matrix NLS equation is also
integrable~\cite{Tsuchida1}.
It is worthy to search other integrable models for higher spin case.
We will discuss such possibilities in a separate paper.

We also mention another reduction:
\begin{equation}
\label{reduction2}
Q=\left(\hspace*{-1mm}\begin{array}{cc}
q_1 \hspace*{-1mm}&\hspace*{-1mm} q_2 \\
0   \hspace*{-1mm}&\hspace*{-1mm} 0
\end{array}
\hspace*{-1mm}\right).
\end{equation}
This gives rise to intensity-coupled NLS equations known as
the Manakov model~\cite{Manakov}:
\begin{eqnarray}
\i \partial _t q_1 + \partial _{x}^2 q _1 + 2 (|q_1|^2+|q_2|^2) q_1 \eq 0,
\nonumber\\
\label{Manakov}
\i \partial _t q_2 + \partial _{x}^2 q _2 + 2 (|q_1|^2+|q_2|^2) q_2 \eq 0,
\end{eqnarray}
which are used to describe the interaction among the modes in nonlinear
optics, e.g., in the case of birefringent and other two-mode fibers~\cite{Rad}.
The extension of the Manakov model (\ref{Manakov})
to the general $m$-component case is straightforward.
Recently, $N$-soliton collisions in this
model have been analyzed in detail~\cite{Tsuchida}.
It should be noted that by rescaling these two fields
$q_1$, $q_2$ appropriately, one can introduce the coupling coefficients
in the nonlinear terms. However, possible choices of the coefficients
that keep the integrability of the system are highly restricted
according to the Painlev\'e analysis~\cite{Saha}.
For example, one can apply the integrable equations
(\ref{Manakov}) to a binary BEC~\cite{binaryBEC} only if both the
inter-species coupling and intra-species coupling coincide. Therefore,
in the Manakov model (\ref{Manakov}),
two (or more in the general case) components
are coupled equally with each other. 
On the other hand, in our spinor model (\ref{stdeq-1}),
it is obvious that 0 component and $\pm1$ components play different roles;
this difference will lead to new types of the soliton solutions and
the unique spin dynamics.

In the following, we shall exploit the exact solution technique which enables
us to solve the initial value problem of the coupled system (\ref{NLS}) with
(\ref{reduction}) and obtain the exact soliton solutions.

\section{\label{sec:ISM}Inverse Scattering Method}
In this section, some useful results obtained by the inverse
scattering method (ISM) are briefly summarized.
We derive an explicit formula for the soliton solution
of the $2 \times 2$ matrix version of NLS equation
(\ref{NLS}) with eq.(\ref{reduction})
by considering a reduction of a formula in ref. 30.

Under the vanishing boundary conditions,
we apply the ISM to the nonlinear time evolution equation (\ref{NLS})
associated with the generalized Zakharov--Shabat eigenvalue problem:
\begin{eqnarray}
\partial_x \left(\hspace*{-1mm}
\begin{array}{l}
\Psi_{\rm I}\\
\Psi_{\rm II}
\end{array}\hspace*{-1mm}
\right)
=\frac{1}{2}\left(\hspace*{-1mm}
\begin{array}{cc}
k^*I \hspace*{-1mm}&\hspace*{-1mm} 2Q \\
-2Q  \hspace*{-1mm}&\hspace*{-1mm} -k^*I
\end{array}\hspace*{-1mm}
\right)
\left(\hspace*{-1mm}
\begin{array}{l}
\Psi_{\rm I}\\
\Psi_{\rm II}
\end{array}\hspace*{-1mm}
\right).
\end{eqnarray}
Here $\Psi_{\rm I}$ and $\Psi_{\rm II}$ take their values in
$2\times 2$ matrices.
The complex number $k$ is the spectral parameter.
$I$ is the $2\times 2$ unit matrix.
The $2\times 2$ matrix $Q$ plays a role as a potential function in this
linear system.

The general $N$-soliton solution of eq.(\ref{NLS}) with eq.(\ref{reduction})
is expressed as~\cite{Tsuchida1}
\begin{eqnarray}
\label{N soliton}
Q(x,t)=(\,\underbrace{I\,I\,\cdots I}_N\,)S^{-1}
\left(\hspace*{-1mm}
\begin{array}{c}
\Pi_1 \e^{\chi_1} \\
\Pi_2 \e^{\chi_2} \\
\vdots \\
\Pi_N \e^{\chi_N}
\end{array}\hspace*{-1mm}
\right),
\end{eqnarray}
where the $2N\times 2N$ matrix $S$ is given by
\begin{eqnarray}
\label{S matrix}
S_{ij}=\delta_{ij}I+\sum_{l=1}^N
       \frac{\Pi_i\cdot\Pi_l^\dagger}{(k_i+k_l^*)(k_j+k_l^*)}
       \e^{\chi_i+\chi_l^*},\\
       1\le i,j \le N.\nonumber
\end{eqnarray}
Here we have introduced the following parameterizations:
\begin{eqnarray}
\Pi_j\eq \left(\hspace*{-1mm}\begin{array}{cc}
\beta_j \hspace*{-1mm}&\hspace*{-1mm}\alpha_j\\
\alpha_j\hspace*{-1mm}&\hspace*{-1mm}\gamma_j
\end{array}\hspace*{-1mm}\right),
\nonumber\\
\chi_j\eequiv\chi_j(x,t)=k_jx+\i k_j^2t-\epsilon_j.
\end{eqnarray}
We explain notations and their significance.
The $2\times2$ matrices $\Pi_j$ normalized to unity in a sense of the 
square norm,
\begin{equation}
||\Pi_j||_2\equiv \sqrt{2|\alpha_j|^2+|\beta_j|^2+|\gamma_j|^2}=1,
\end{equation}
must take the same form as $Q$ from their definition.
We call them ``polarization matrices" as usual in the Manakov model.
In the spinor model, the polarization matrices determine both 
the populations of three components $\{1,\,0,\,-1\}$ within each soliton
and the relative phases between them.
The complex constants $k_j$ denote discrete eigenvalues,
each of which determines a bound state by
the potential $Q$. $\epsilon_j$ are real constants which can be used to
tune the initial displacements of solitons.
It is worth noting that all $x$ and $t$ dependence is only through
the variables $\chi_j(x,t)$.
As we shall see in section~\ref{sec:OSS}, the real part of $\chi_j(x,t)$
represents the coordinate for observing soliton-$j$'s envelope while
the imaginary part of it represents the coordinate for observing
soliton-$j$'s carrier waves.

Equation (\ref{NLS}) is a completely integrable system
whose initial value problems can be solved
via, for example, the ISM.
The existence of the $\mathbf r$-matrix for this system 
guarantees the existence of an infinite number of conservation laws
which restrict the dynamics of the system in an essential way.
Here we show explicit forms of some conserved quantities,
i.e., total number, total spin, total momentum and total energy.
\begin{eqnarray}
\label{total number}
\hspace*{-5mm}&&\hspace*{-5mm}
\mbox{\bf total number:}\quad 
N_{\rm T}= \int\d x \,n(x,t)\,;\\
\label{lnd}
\hspace*{-5mm}&&\hspace*{2mm}
n(x,t)={\bm\Phi}^\dagger\cdot{\bm\Phi}
={\rm tr}\{Q^\dagger Q\}.\\
\label{total spin}
\hspace*{-5mm}&&\hspace*{-5mm}
\mbox{\bf total spin:}\quad 
{\mathbf F}_{\rm T}= \int\d x \,{\mathbf f}(x,t)\,;\\
\label{lsd}
\hspace*{-5mm}&&\hspace*{2mm}
{\mathbf f}(x,t)={\bm\Phi}^\dagger\!\cdot\!
{\bm f}\!\cdot\!{\bm\Phi}
={\rm tr}\{Q^{\dagger} {\bm \sigma} Q\}.\\
\label{tot mom}
\hspace*{-5mm}&&\hspace*{-5mm}
\mbox{\bf total momentum:}\quad 
P_{\rm T}= \int\d x \,p(x,t)\,; \\
\label{lmom}
\hspace*{-5mm}&&\hspace*{2mm}
p(x,t)= -\i\hbar{\bm\Phi}^\dagger\cdot\partial_x{\bm\Phi}
= -\i\hbar \cdot{\rm tr}\{Q^\dagger Q_x\}.\\
\label{tot ene}
\hspace*{-5mm}&&\hspace*{-5mm}
\mbox{\bf total energy:}\quad 
E_{\rm T}= \int\d x \,e(x,t)\,;\\
\hspace*{-5mm}&&\hspace*{2mm}
e(x,t)=
\frac{\hbar^2}{2m}\partial_x{\bm\Phi}^\dagger\cdot\partial_x{\bm\Phi}
-\frac{c}{2}\left[n(x,t)^2+{\mathbf f}(x,t)^2 \right]\nonumber\\
\label{lene}
\hspace*{-5mm}&&\hspace*{12mm}
=\mbox{}c\cdot{\rm tr}\{Q^\dagger_{x}Q_{x}-Q^\dagger QQ^\dagger Q\}.
\end{eqnarray}
Here tr$\{\cdot\}$ denotes the matrix trace.
In eq.(\ref{lsd}), the expression (\ref{spin1 mtx}) and 
the Pauli matrices $\bm{\sigma}=(\sigma^x,\sigma^y,\sigma^z)^T$ are used.
These conservation laws can be proved by a direct calculation.
For a complete set of the conserved densities and
the recursion formula for them, see discussion in ref. \citen{Tsuchida1}. 

\section{\label{sec:OSS}Spin State of One-Soliton Solution}
In this section, we discuss one-soliton solution
and the classification of its spin states in the spinor model.
If we set $N=1$ in the formula (\ref{N soliton}),
we obtain the one-soliton solution:
\begin{equation}
\label{OSS}
Q
=\frac{\e^{\chi}}{{\rm det}S}\left(\begin{array}{cc}
\hspace*{-1mm}
\beta +\gamma^*\e^{2\chi_{\rm R}+\rho}{\rm det}\Pi 
\hspace*{-1mm}&\hspace*{-1mm}
\alpha-\alpha^*\e^{2\chi_{\rm R}+\rho}{\rm det}\Pi
\hspace*{-1mm} \\
\hspace*{-1mm}
\alpha-\alpha^*\e^{2\chi_{\rm R}+\rho}{\rm det}\Pi
\hspace*{-1mm}&\hspace*{-1mm}
\gamma+\beta^*\e^{2\chi_{\rm R}+\rho}{\rm det}\Pi
\hspace*{-1mm}
\end{array}\right),
\end{equation}
where
\begin{eqnarray}
\label{det}
&&{\rm det}S=1+\e^{2\chi_{\rm R}+\rho} 
+\e^{4\chi_{\rm R}+2\rho} |{\rm det}\Pi|^2,\\
\label{param1}
&&\e^{\rho/2}\equiv\frac{1}{2k_{\rm R}}, \quad
\Pi\equiv\left(\hspace*{-1mm}\begin{array}{cc}
\beta \hspace*{-1mm}&\hspace*{-1mm} \alpha \\
\alpha \hspace*{-1mm}&\hspace*{-1mm} \gamma
\end{array}\hspace*{-1mm}\right),\\
\label{param2}
&&\chi_{\rm R}\equiv\chi_{\rm R}(x,t)= k_{\rm R}(x-2k_{\rm I}t)-\epsilon,\\
\label{param3}
&&\chi_{\rm I}\equiv\chi_{\rm I}(x,t)= k_{\rm I}x+(k_{\rm R}^2-k_{\rm I}^2)t.
\end{eqnarray}
We have omitted the subscripts of the soliton number.
Here and henceforth, the subscripts R and I denote real and imaginary parts,
respectively.
Throughout this section, we set $k_{\rm R}>0$ without loss of generality.
In a different form:
\begin{eqnarray}
\label{OSS2}
Q=2k_{\rm R}\frac{\Pi\,\e^{-(\chi_{\rm R}+\rho/2)}
+\left(\sigma^y\Pi^\dagger\sigma^y\right)\e^{\chi_{\rm R}+\rho/2}{\rm det}\Pi}
{\e^{-(2\chi_{\rm R}+\rho)}+1
+\e^{2\chi_{\rm R}+\rho} |{\rm det}\Pi|^2}\e^{\smalli \chi_{\rm I}},
\end{eqnarray}
we can make out the significance
of each parameter/coordinate as follows,
\begin{eqnarray*}
&&\hspace*{-4mm}
 \Pi:\,\mbox{polarization matrix of soliton}\\
&&\hspace*{-4mm}
 k_{\rm R}:\,\mbox{amplitude of soliton}\\
&&\hspace*{-4mm}
 2k_{\rm I}:\,\mbox{velocity of soliton's envelope}\\
&&\hspace*{-4mm}
 \chi_{\rm R}:\,\mbox{coordinate for observing soliton's envelope}\\
&&\hspace*{-4mm}
 \chi_{\rm I}:\,\mbox{coordinate for observing soliton's carrier waves}.
\end{eqnarray*}
A few comments will be made on those points pertinent to this list.
We use the term ``amplitude" to indicate the peak(s) height
of soliton's envelope.
Actual amplitude should be represented as $k_{\rm R}$ multiplied by a factor
from 1 to $\sqrt{2}$ which is determined by the type of polarization matrices.
The explicit form will be shown later.
As mentioned before, soliton's motion depends on both $x$ and $t$ via
variables $\chi_{\rm R}$ and $\chi_{\rm I}$,
from which we can see the meaning of velocity of soliton.

From a total spin conservation, one-soliton solution
can be classified by the spin states.
We shall show that the only two spin states are allowable,
i.e., ${\bf F}_{\rm T}=(0,0,0)^T$ for ${\rm det}\Pi\ne 0$, and 
$|{\bf F}_{\rm T}|=N_{\rm T}$ for ${\rm det}\Pi=0$.

Substituting eqs.(\ref{OSS})--(\ref{param3}) into eq.(\ref{lsd}),
we obtain the local spin density of the one-soliton solution:
\begin{equation}
\label{spin density}
{\bf f}(x,t)=\frac{\e^{2\chi_{\rm R}}}{({\rm det}S)^2}
\left(1-\e^{4\chi_{\rm R}+2\rho}|{\rm det}\Pi|^2\right)
{\rm tr}\{\Pi^{\dagger} {\bm \sigma} \Pi\}.
\end{equation}
We also give the explicit form of the number density:
\begin{equation}
\label{number density}
n(x,t)=\frac{\e^{2\chi_{\rm R}}}{({\rm det}S)^2}
\left\{1+\left(4\e^{2\chi_{\rm R}+\rho}
+\e^{4\chi_{\rm R}+2\rho}\right)|{\rm det}\Pi|^2\right\}.
\end{equation}

To clarify the physical meaning of $\det\Pi$,
we define here another important local density as
\begin{equation}
\label{singlet pair}
\Theta(x,t) \equiv\Phi_0^2-2\Phi_1\Phi_{-1}=-2\det Q.
\end{equation}
This quantity measures the formation of singlet pairs.
Note that these ``pairs" are distinguished from Cooper pairs of electrons
or those of ${}^3$He owing to the different statistical properties of 
ingredient particles~\cite{Koashi,Ueda}.
Since $\Theta(x,t)$ does not contribute to the magnetization of the soliton,
it is invariant under any spin rotation~\cite{Ciobanu}.
As far as ground state properties are concerned,
it is not necessary to introduce $\Theta(x,t)$ for a system of spin-1 bosons,
while a counterpart to eq.(\ref{singlet pair}) plays a crucial role for spin-2
case~\cite{Koashi,Ciobanu,Ueda}.
However, as we shall show later, it is useful to characterize solitons within
energy degenerated states.

In the case of the one-soliton solution (\ref{OSS}),
the singlet pair density is proportional to the determinant
of the polarization matrix $\Pi$,
\begin{equation}
\label{singlet}
\Theta(x,t)=-2\frac{\e^{2\chi}}{\det S}\det\Pi.
\end{equation}
This suggests that $\det\Pi$ represents the magnitude of the singlet pairs
locally.
For the general case, however, this singlet pair density can vary after each
collision of solitons and is not the conserved density.
The detail will be discussed at the end of this section and also
in \S \ref{subsec:p-f}.  
Now, we classify spin states of the one-soliton solution
into two distinct cases depending on the values of $\det\Pi$.

\subsection{Ferromagnetic state}
Let ${\rm det}\Pi=0$, then eq.(\ref{OSS}) becomes a simple form:
\begin{equation}
\label{OSSfero}
Q=k_{\rm R}{\rm sech}(\chi_{\rm R}+\rho/2)\Pi\e^{\smalli\chi_{\rm I}}.
\end{equation}
Now all of $m_F=0,\,\pm1$ components share the same wave function.
Their distribution in the internal state reflects the elements of
the polarization matrix $\Pi$ directly. One can clearly see
the meaning of each parameter listed above.
By definition, the singlet pair density (\ref{singlet}) vanishes
everywhere. Thus, this type of soliton belongs to the ferromagnetic state
and will be referred to as a ferromagnetic soliton.
The total number of atoms is obtained by integrating
eq.(\ref{number density}) as $N_{\rm T}=2k_{\rm R}$.
The total spin (\ref{total spin}) becomes 
\begin{eqnarray}
\label{OSspin}
{\bf F}_{\rm T}=2k_{\rm R}
\left(\hspace*{-1mm}\begin{array}{c}
2\,{\rm Re}\!\left\{\alpha^*(\beta+\gamma)\right\}\\
-2\,{\rm Im}\!\left\{\alpha^*(\beta-\gamma)\right\}\\
|\beta|^2-|\gamma|^2
\end{array}\hspace*{-1mm}\right),\,|{\bf F}_{\rm T}|=N_{\rm T},
\end{eqnarray}
which is connected to ${\bf F}'_{\rm T}=2k_{\rm R}(0,0,1)^T$
through a gauge transformation and a spin rotation.

Next, we calculate the total momentum and the total energy of
the ferromagnetic soliton.
Substituting eq.(\ref{OSSfero}) into eqs.(\ref{tot mom}), (\ref{tot ene}),
and using ${\rm det}\Pi=0$, we obtain
\begin{eqnarray}
P_{\rm T}^{\rm f}\eq-\i\hbar k_{\rm R}^2\int\d x 
\left\{\i k_{\rm I}-k_{\rm R}\tanh(\chi_{\rm R}+\rho/2)\right\}\nonumber\\
\espace\qquad\qquad{}\times{\rm sech}^2(\chi_{\rm R}+\rho/2)\nonumber\\
\eq N_{\rm T}\hbar k_{\rm I},\\
E_{\rm T}^{\rm f}\eq c k_{\rm R}^2\int\d x 
\left[k_{\rm R}^2\left\{2\tanh^2(\chi_{\rm R}+\rho/2)-1\right\}
+k_{\rm I}^2\right]\nonumber\\
\espace\qquad\qquad{}\times{\rm sech}^2(\chi_{\rm R}+\rho/2)\nonumber\\
\eq N_{\rm T} c \left( k_{\rm I}^2-\frac{k_{\rm R}^2}{3}\right),
\end{eqnarray}
respectively.

\subsection{Polar state}
If ${\rm det}\Pi\ne 0$, the local spin density has one node,
i.e., ${\bf f}(x_0,t)=0$ at a point:
\begin{equation}
x_0= 2k_{\rm I}t+\frac{1}{2k_{\rm R}}\left(
\ln{\frac{4k_{\rm R}^2}{|{\rm det}\Pi|}}+2\epsilon\right),
\end{equation}
for each moment $t$.
Setting $x^\prime=x-x_0$ and $A^{-1}\equiv 2|\det\Pi|$, we get
\begin{equation}
\label{polar-s}
{\bf f}(x^\prime)=-\frac{
4k_{\rm R}^2A\sinh(2k_{\rm R}x^\prime)}
{\big[\cosh(2k_{\rm R}x^\prime)+A\big]^2}
\left(\hspace*{-1mm}\begin{array}{ccc}
2\,{\rm Re}\!\left\{\alpha^*(\beta+\gamma)\right\}\\
-2\,{\rm Im}\!\left\{\alpha^*(\beta-\gamma)\right\}\\
|\beta|^2-|\gamma|^2
\end{array}\hspace*{-1mm}\right).
\end{equation}
Since each component of the local spin density is an odd function
of $x^\prime$, its average value is zero,
\begin{equation}
{\bf F}_{\rm T}=\int \d x^\prime\,
{\bf f}(x^\prime)
=\left(\hspace*{-1mm}\begin{array}{ccc}
0\\0\\0
\end{array}\hspace*{-1mm}\right).
\end{equation}
This implies that this type of soliton, on the average,
belongs to the polar state~\cite{Ho}. 
Let us also rewrite the number density (\ref{number density}) as
\begin{equation}
\label{polar-num}
n(x^\prime)=\frac{4k_{\rm R}^2
\left[A \cosh{2k_{\rm R}x^\prime}+1\right]}
{\left[\cosh{2k_{\rm R}x^\prime}+A\right]^2}.
\end{equation}
To elaborate on this type of soliton, we further divide into two cases.\\

\noindent
(i) $A^{-1}= 2|\det\Pi|=1$
($\alpha\beta^*+\alpha^*\gamma$=0).

Under this constraint, we find the local spin (\ref{polar-s}) itself
vanishes everywhere.
Solitons in this state possess the symmetry of polar state locally.
We, therefore, refer to only those solitons as polar solitons in this paper.
Considering eq.(\ref{OSS2}) with the above condition, we recover a normal
sech-type soliton solution:
\begin{equation}
\label{OSSpol}
Q=\sqrt{2}k_{\rm R}{\rm sech}(k_{\rm R}x^\prime)\Pi\e^{\smalli\chi_{\rm I}}.
\end{equation}
Note that the amplitude of soliton is different from that of the
ferromagnetic soliton, which leads to a relation between the total number
and the spectral parameter as $N_{\rm T}=4k_{\rm R}$.
The total momentum and the total energy are given by
$P_{\rm T}^{\rm p}=N_{\rm T}\hbar k_{\rm I}$,
$E_{\rm T}^{\rm p}=N_{\rm T} c \left( k_{\rm I}^2-k_{\rm R}^2/3\right)$,
respectively.
The difference between ferromagnetic soliton energy and polar soliton energy
with the same number of atoms $N_{\rm T}$ is
\begin{equation}
E_{\rm T}^{\rm f}-E_{\rm T}^{\rm p}=-\frac{N_{\rm T}^3 c}{16}<0,
\end{equation}
which is a natural consequence of the ferromagnetic interaction, i.e., $c_2<0$.
\\

\noindent
(ii) $A^{-1}= 2|\det\Pi|< 1$.

\begin{figure}[bp]
\begin{center}
    \includegraphics[height=.2\textheight]{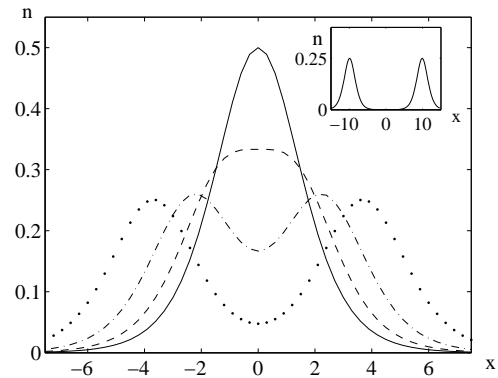}
\end{center}
  \caption{The density profiles of eq.(\ref{polar-num}).
  We set $k_{\rm R}=0.5$, and $A=1$ (solid line),
  2 (dashed line), 5 (dash-dot line), 20 (dotted line).
  The inset shows a split soliton for $A=10^4$, 
  consisting of two ferromagnetic like solitons
  with the same velocity.}
  \label{fig:split}
\end{figure}

\begin{figure}[tbp]
\begin{center}
    \includegraphics[height=.2\textheight]{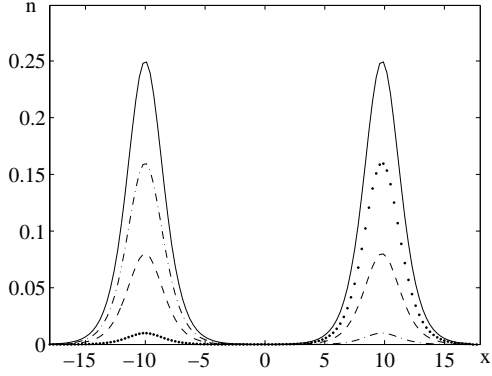}
\end{center}
  \caption{The density profiles of eq.(\ref{polar-num}) (solid line)
  for $k_{\rm R}=0.5$ and $A=10^4$, and the three components,
  $m_F=0$ (dashed line), $m_F=1$ (dotted line)
  and $m_F=-1$ (dash-dot line) are shown simultaneously.}
  \label{fig:split2}
\end{figure}

In this case, the local spin retains nonzero value,
although the average spin amounts to be zero.
The density profile has a complicated form as eq.(\ref{polar-num}).
Note that when $A>2$, a peak of the density splits
into two (Fig.\ref{fig:split}) due to different density profiles of
$m_F=0,\,\pm1$ components.
For a large value of $A$, viz., when $\det\Pi$ gets close to zero,
such twin peaks separate away.
In consequence, they behave as if a pair of two distinct ferromagnetic solitons
with antiparallel spins,
traveling in parallel with the same velocity and the amplitudes half as much as
that of the polar soliton ($A=1$) in the density profile
(see the inset of Fig.\ref{fig:split} and Fig.\ref{fig:split2}).
Hence, solitons of this type will be referred to as split solitons.
The total number is the same as (i) case, $N_{\rm T}=4k_{\rm R}$,
which can be shown by changing a variable $y=\tanh(k_{\rm R}x^\prime)$
and evaluating a definite integral:
\begin{eqnarray}
\label{def int}
\int^{1}_{-1}\d y\,\frac{(A+1)+y^2(A-1)}
{\left[(A+1)-y^2(A-1)\right]^2}=1.
\end{eqnarray}

By use of the similar definite integrals, the total momentum and the total energy
are shown to be the same values as those in the case (i):
$P_{\rm T}^{\rm s}=N_{\rm T}\hbar k_{\rm I}$,
$E_{\rm T}^{\rm s}=N_{\rm T} c \left( k_{\rm I}^2-k_{\rm R}^2/3\right)$.
This degeneracy is ascribed to the integrable condition for the coupling
constants, i.e., $c_0=c_2$.

Comparing case (i) with case (ii), we find that a variety of dissimilar
shaped solitons are degenerated in the polar state.
To characterize them, we can use, instead of $A$ itself,
a physical quantity defined as
\begin{eqnarray}
\label{tot sing}
\mathcal{S}\eequiv\int\d x|\Theta(x,\,t)|\nonumber\\
\eq N_{\rm T}\frac{2 \tan^{-1}\left(\sqrt{\frac{A-1}{A+1}}\right)}{\sqrt{A^2-1}},
\end{eqnarray}
which is a monotone decreasing function of $A\in[1,\,\infty)$;
the maximum value, $N_{\rm T}$, at $A=1$ (polar soliton)
and limiting to 0 in $A\to\infty$ (ferromagnetic soliton).
In this sense, $\mathcal{S}$ has the meaning as
the ``total singlet pairs" of the whole system.
As noted above, $\mathcal{S}$ is not the conserved quantity in general ($N\ge2$);
all the conserved densities should be expressed by the matrix trace of
products of $Q^\dagger$, $Q$ and their derivatives~\cite{Tsuchida1}
as eqs.(\ref{lnd}), (\ref{lsd}), (\ref{lmom}), and (\ref{lene})
while $|\Theta(x,\,t)|$ is not.
Nevertheless, $\mathcal{S}$ can be used to label solitons in the polar state
because it dose not change in the meanwhile prior to the subsequent collision.

\section{\label{sec:switch}Two-Soliton Collision and Scattering Law}
In this section, we analyze two-soliton collisions
in the spinor model.
The two-soliton solution can be obtained by setting $N=2$ in
eq.(\ref{N soliton}). The derivation is straightforward but
rather lengthy. We give a general formula of the two-soliton solution
in Appendix and, here, compute asymptotic forms of specific two-soliton
solutions as $t\to \mp \infty$, which define the collision laws
of two-soliton in the spinor model.
For simplicity, we restrict the spectral parameters to regions:
\begin{eqnarray}
k_{1{\rm R}}\hspace{-1mm}&>0,\quad k_{2{\rm R}}\hspace{-1mm}&<0,\nonumber\\
\label{head on}
k_{1{\rm I}}\hspace{-1mm}&<0,\quad k_{2{\rm I}}\hspace{-1mm}&>0.
\end{eqnarray}
Under the conditions, we calculate the asymptotic forms in the final state
($t\to\infty$) from those in the initial state ($t\to -\infty$).
Since each soliton's envelope is located around $x\simeq2k_{j{\rm I}}t$,
soliton-1 and soliton-2 are initially isolated at $x\to\pm\infty$, 
and then, travel to the opposite direction at a velocity
of $2k_{1{\rm I}}$ and $2k_{2{\rm I}}$, respectively.
After a head-on collision, they pass through without changing their
velocities and arrive at $x\to\mp\infty$ in the final state.
Collisional effects appear not only as usual phase shifts of solitons
but also as a rotation of their polarization.

According to the classification of one-soliton solutions in the previous 
section, we choose the following three cases:
(1) polar-polar solitons collision,
(2) polar-ferromagnetic solitons collision, and
(3) ferromagnetic-ferromagnetic solitons collision.
As we shall see later, the polar soliton dose not affect the polarization
of the other solitons apart from the total phase factor.
On the other hand, ferromagnetic solitons can `rotate' their partners'
polarization, which allows for switching among the internal states.

\subsection{\textit{Polar-polar solitons collision}}
We first deal with a collision between two polar solitons defined by
$k_j$ and $\Pi_j$ ($j=1,\,2$) with the conditions 
(\ref{head on}) and $|\det\Pi_1|=|\det\Pi_2|=1/2$.
In the asymptotic regions, we can consider each soliton separately.
Thus, the initial state is given by the sum of two polar solitons as
\begin{eqnarray}
\label{TSS-in}
Q\simeq Q_1^{\rm in}+Q_2^{\rm in},
\end{eqnarray}
where the asymptotic form of soliton-$j$ ($j=1,\,2$) is 
\begin{equation}
\label{TSSpol-in}
Q_j^{\rm in}=
\sqrt{2}k_{j{\rm R}}{\rm sech}(\chi_{j{\rm R}}+\rho_j/2)\Pi_j
\e^{\smalli\chi_{j{\rm I}}}.
\end{equation}
These can be proved by taking
the limit $\chi_{2{\rm R}}\to -\infty$ with keeping $\chi_{1{\rm R}}$ finite
and, vice versa, $\chi_{1{\rm R}}\to -\infty$ with $\chi_{2{\rm R}}$ fixed.
Phase factors which come from the values of $|\det\Pi_j|$
are absorbed by the arbitrary constants $\epsilon_j$ inside $\chi_{j{\rm R}}$.
In the final state, the opposite limit $\chi_{2{\rm R}}\to \infty$
with keeping $\chi_{1{\rm R}}$ finite and $\chi_{1{\rm R}}\to \infty$
with $|\chi_{2{\rm R}}|<\infty$  yield 
\begin{eqnarray}
\label{TSS-fin}
Q\simeq Q_1^{\rm fin}+Q_2^{\rm fin},
\end{eqnarray}
where 
\begin{equation}
\label{TSSpol-fin}
Q_j^{\rm fin}=
\sqrt{2}k_{j{\rm R}}{\rm sech}\left(\chi_{j{\rm R}}+\rho_j/2+r\right)
\Pi_j\e^{\smalli(\chi_{j{\rm I}}+\sigma_j)},
\end{equation}
with
\begin{eqnarray}
\label{r}
r\eq
2\ln\left|\frac{k_1-k_2}{k_1+k_2^*}\right|,\\
\label{phase shift2}
\sigma_1\eq 2 \arg\left(\frac{k_1-k_2}{k_1+k_2^*}\right),\,\,
\sigma_2= 2 \arg\left(\frac{k_2-k_1}{k_2+k_1^*}\right).
\end{eqnarray}

Equations (\ref{TSSpol-in}) and (\ref{TSSpol-fin}) are the same form as polar
one-soliton solution (\ref{OSSpol}).
Collisional effects appear only in the position shift (\ref{r})
and the phase shifts (\ref{phase shift2}).
Thus, the partial number $N_j$, spin $F_j$, momentum $P_j$, and energy $E_j$
are defined for the asymptotic form of soliton-$j$ and calculated
in the same manner as the previous section. The integrals of motion are
represented by the sum of those quantities for each soliton.
Moreover, we can prove $N_j=4|k_{j{\rm R}}|$,
$|F_j|=0$, $P_j=N_j\hbar k_{j{\rm I}}$, and
$E_j=N_jc(k_{j{\rm I}}^2-k_{j{\rm R}}^2/3)$ themselves
are conserved through the collision.
In this sense, the polar-polar collision is basically the same as that of
the single-component NLS equation.

\subsection{\label{subsec:p-f}\textit{Polar-ferromagnetic solitons collision}}
Under the condition (\ref{head on}),
we set soliton 1 to be polar soliton ($|\det\Pi_1|=1/2$),
and soliton 2 to be ferromagnetic soliton ($|\det\Pi_2|=0$).
Then, the initial state is represented by eq.(\ref{TSS-in}) with
\begin{eqnarray}
\label{TSSpf-1in}
Q_1^{\rm in}\eq
\sqrt{2}k_{1{\rm R}}{\rm sech}(\chi_{1{\rm R}}+\rho_1/2)\Pi_1
\e^{\smalli\chi_{1{\rm I}}},\nonumber\\
\label{TSSpf-2in}
Q_2^{\rm in}\eq
k_{2{\rm R}}{\rm sech}(\chi_{2{\rm R}}+\rho_2/2)\Pi_2
\e^{\smalli\chi_{2{\rm I}}}.
\end{eqnarray}
The final state is given by eq.(\ref{TSS-fin}) with
\begin{eqnarray}
\label{TSSpf-1fin}
Q_1^{\rm fin}\eq
2 k_{1{\rm R}}\e^{\smalli\chi_{1{\rm I}}}
\nonumber\\
\espace{}\hspace*{-5mm}\times
\frac{\tilde{\Pi}_1\e^{-(\chi_{1{\rm R}}+\rho_1/2+\delta)}+
\big(\sigma_y\tilde{\Pi}_1^\dagger\sigma_y\big){\rm det}\tilde{\Pi}_1
\e^{\chi_{1{\rm R}}+\rho_1/2+\delta}}
{\e^{-(2\chi_{1{\rm R}}+\rho_1+2\delta)}+1
+\e^{2\chi_{1{\rm R}}+\rho_1+2\delta }|\det\tilde{\Pi}_1|^2}
,\nonumber\\
\label{TSSpf-2fin}
Q_2^{\rm fin}\eq
k_{2{\rm R}}{\rm sech}\left(\chi_{2{\rm R}}+\rho_2/2+r\right)
\Pi_2\e^{\smalli(\chi_{2{\rm I}}+\sigma_2)}.
\end{eqnarray}
Here we have defined
\begin{eqnarray}
\label{delta}
\e^{2\delta}\eq\left|\frac{k_1-k_2}{k_1+k_2^*}\right|^2\nonumber\\
\espace{}\hspace*{-5mm}\times
\left\{1+\frac{(k_1+k_1^*)^2(k_2+k_2^*)^2}{|k_1+k_2^*|^2|k_1-k_2|^2}
\big|{\rm tr}\big(\Pi_1\Pi_2^\dagger\big)\big|^2\right\},\\
\label{new pi}
\tilde{\Pi}_1\eq
\e^{-\delta}\bigg\{\Pi_1-\frac{k_2+k_2^*}{k_1+k_2^*}
\left(\Pi_1\Pi_2^\dagger\Pi_2+\Pi_2\Pi_2^\dagger\Pi_1\right)\nonumber\\
\espace {}+
\left(\frac{k_2+k_2^*}{k_1+k_2^*}\right)^2{\rm tr}\big(\Pi_1\Pi_2^\dagger
\big)\Pi_2
\bigg\},
\end{eqnarray}
and also used eqs.(\ref{r}), (\ref{phase shift2}).
Normalization of the new polarization matrix (\ref{new pi}) turns to be
unity, $||\tilde{\Pi}_1||_2=1$. The determinant of it becomes
\begin{equation}
\det\tilde{\Pi}_1=\e^{-2\delta}\left(\frac{k_1-k_2}{k_1+k_2^*}\right)^2
\det\Pi_1.
\end{equation}

We can see clearly that the initial polar soliton breaks into a split type
(${\tilde A}_1\equiv (2|\det{\tilde\Pi}_1|)^{-1}>1$)
after the collision with a ferromagnetic one.
Only when $\big|{\rm tr}\big(\Pi_1\Pi_2^\dagger\big)\big|=0$ (in this case, 
the spinor wavefunctions of two initial solitons are orthogonal),
we have ${\tilde A}_1=1$. Then, eqs.(60) are reduced to
\begin{eqnarray}
Q_1^{\rm fin}\eq
\sqrt{2}k_{1{\rm R}}{\rm sech}\left(\chi_{1{\rm R}}+\rho_1/2+r\right)
\Pi_1\e^{\smalli(\chi_{1{\rm I}}+\sigma_1)},\nonumber\\
Q_2^{\rm fin}\eq
k_{2{\rm R}}{\rm sech}\left(\chi_{2{\rm R}}+\rho_2/2+r\right)
\Pi_2\e^{\smalli(\chi_{2{\rm I}}+\sigma_2)}.
\end{eqnarray}
which means that the polar soliton keeps its shape after the collision and
shows no mixing among the internal states except for the total phase shift.
On the other hand, because of the total spin conservation, 
the ferromagnetic soliton always retains its polarization matrix and shows
only the position and phase shifts similar to those of the polar-polar case.

\begin{figure}[tbp]
  \begin{center} 
    \includegraphics[height=.19\textheight]{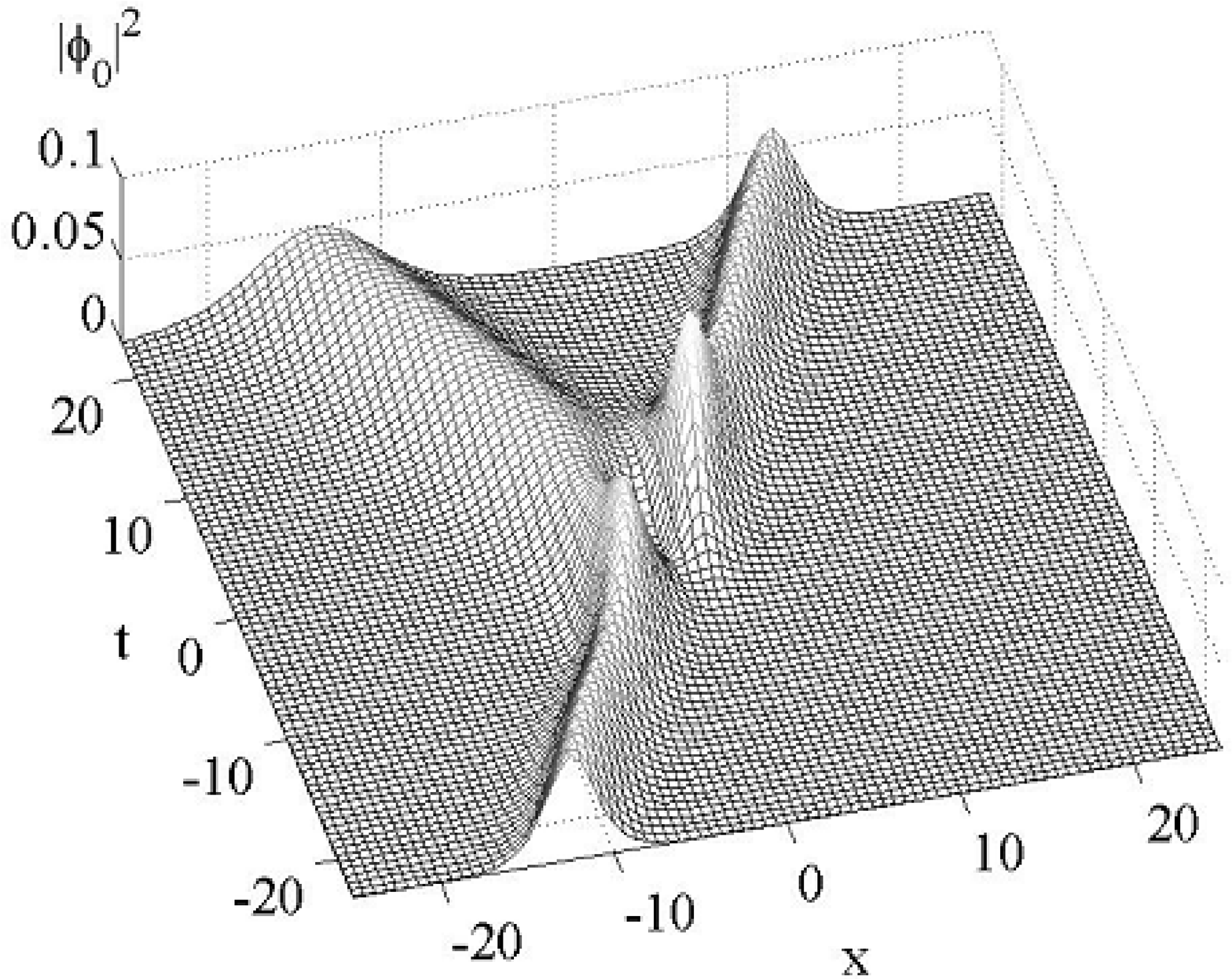}
  \end{center}
  \begin{center}
    \includegraphics[height=.19\textheight]{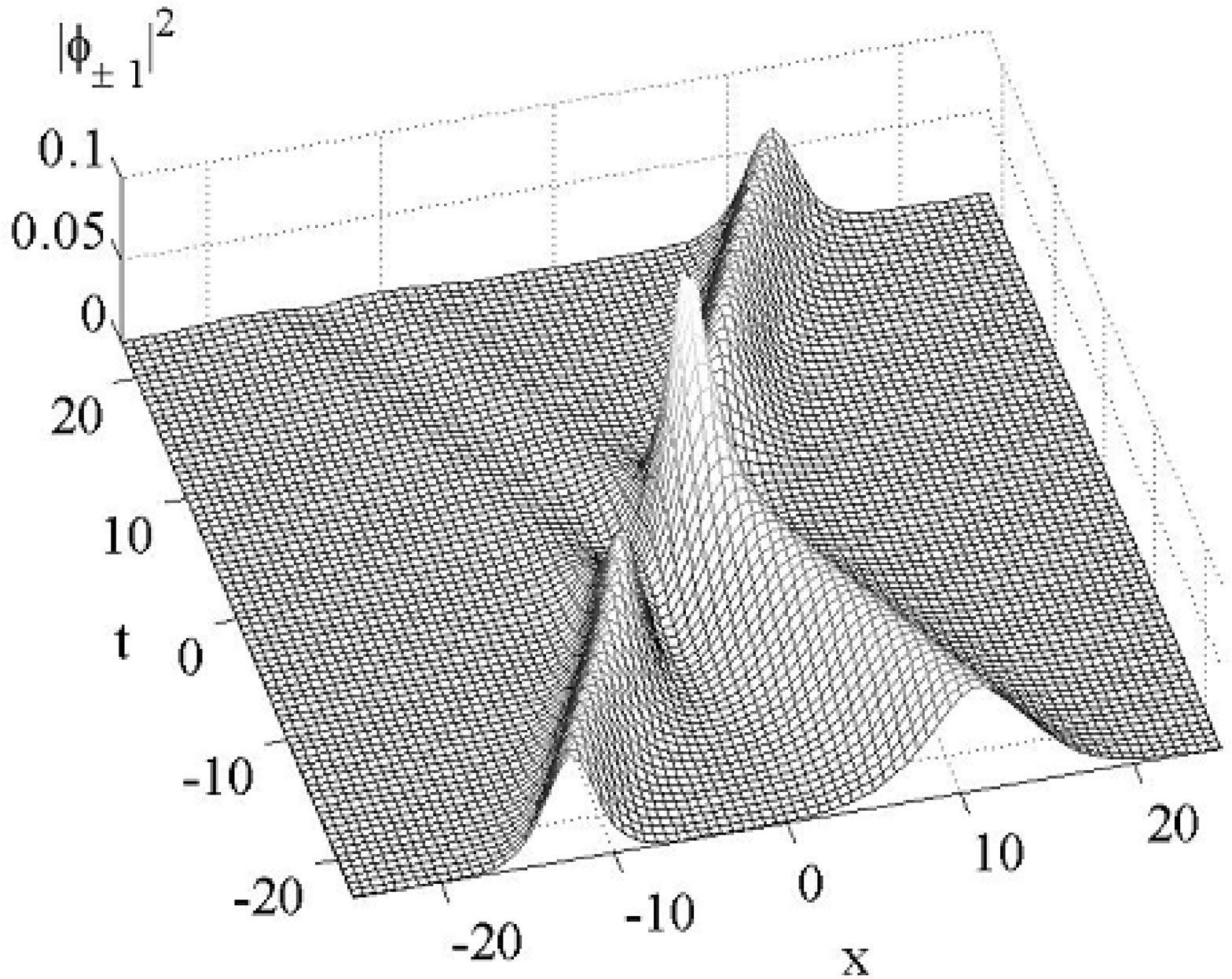}
  \end{center}
  \caption{Density plots of $|\phi_0|^2$ (top) and $|\phi_{\pm 1}|^2$
  (bottom) for a polar-ferromagnetic collision. Soliton 1 (left mover)
  is a polar soliton and soliton 2 (right mover) is a ferromagnetic soliton.
  The parameters used here are $k_1=0.25-0.25\i$, $k_2=-0.5+0.25\i$,
  $\alpha_1=0$, $\beta_1=\gamma_1=1/\sqrt{2}$,
  $\alpha_2=\beta_2=\gamma_2=1/2$.}
  \label{fig:pf_mesh}
  \begin{center}
    \includegraphics[height=.20\textheight]{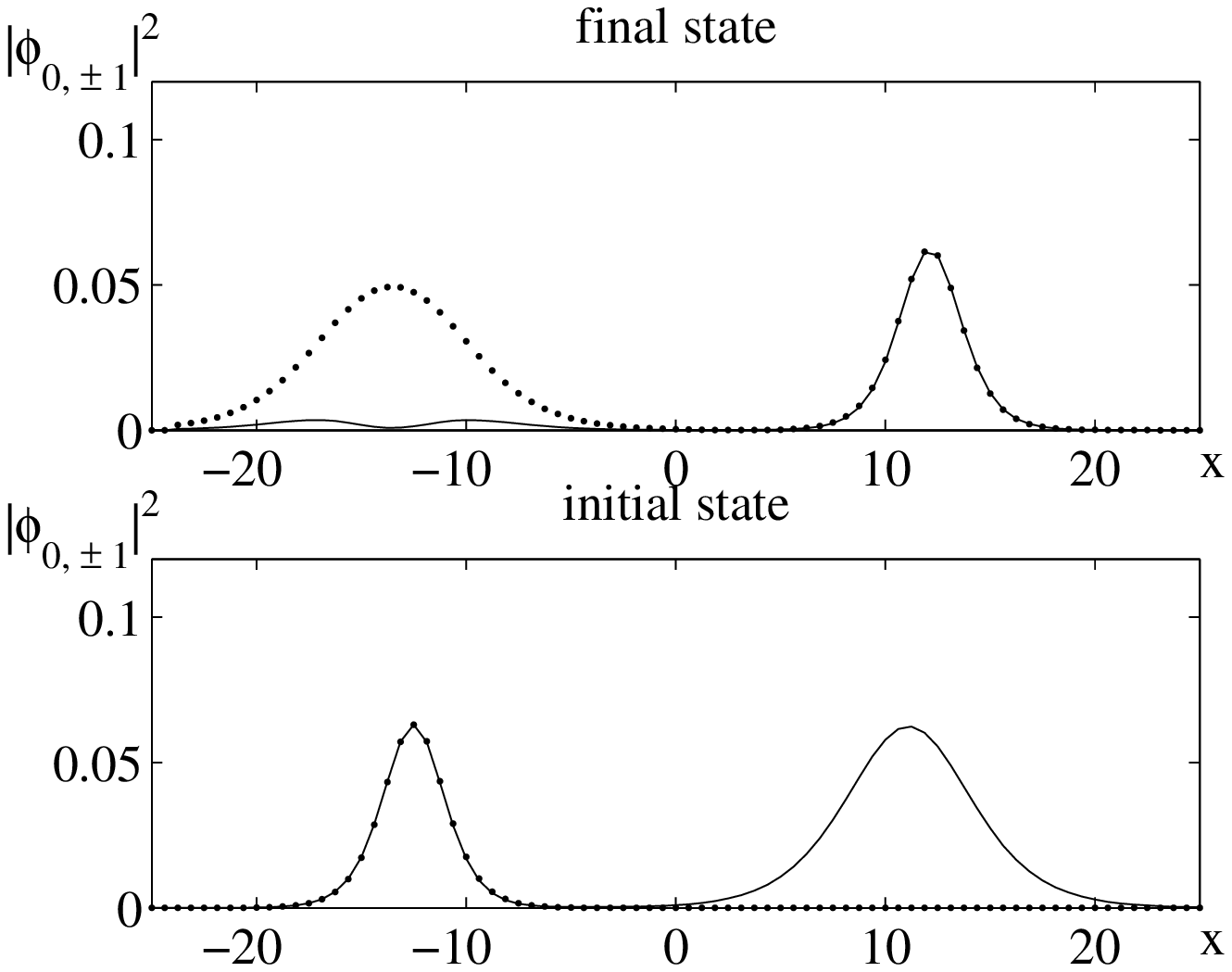}
  \end{center}
  \caption{Density plots of $|\phi_0|^2$ (dots) and $|\phi_{\pm 1}|^2$
  (solid line) in the initial state (bottom) and the final state (top)
  of Fig.\ \ref{fig:pf_mesh}.}
  \label{fig:pf_asympt}
\end{figure}

In Fig.\ \ref{fig:pf_mesh}, we have density plots of a polar-ferromagnetic
collision with the parameters shown in the caption.
These pictures correspond to each component of the exact two-soliton solution
(\ref{2SS}) for one collisional run.
For simplicity, we choose the parameters
to have $|\phi_1|=|\phi_{-1}|$. Figure \ref{fig:pf_asympt} shows the overlap
of $|\phi_0|^2$ and $|\phi_{\pm 1}|^2$ in the initial state (\ref{TSSpf-1in})
and in the final state (\ref{TSSpf-1fin}).
The polar soliton (soliton 1) initially prepared in $m_F={\pm 1}$ are
switched into a soliton with a large population in $m_F=0$ and
the remnant of $m_F={\pm 1}$ after the collision.
Through the collision, the ferromagnetic soliton (soliton 2) plays
only a switcher, showing no mixing in the internal state of itself
outside the collisional region, as clearly seen in eq.(\ref{TSSpf-2fin}). 
In general, this kind of a drastic internal shift of polar soliton is likely
observed for large values of $\big|{\rm tr}\big(\Pi_1\Pi_2^\dagger\big)\big|$
which appears in eqs.(\ref{delta}), (\ref{new pi}).
Although all the conserved quantities such as the number of particles and
the averaged spin of individual solitons are invariant during this type of
collision, the fraction of each component can vary not only in each soliton
level but also in the total after the collision.
This contrasts to an intensity coupled multi-component system
in which the total distribution 
among all components is invariant throughout soliton collisions.

\subsection{\textit{Ferromagnetic-ferromagnetic solitons collision}}

As an instructive example, we already discussed the collision
between two ferromagnetic solitons ($\det \Pi_1=\det\Pi_2=0$) in ref. 23.
In this subsection, we recapitulate it for completeness of this paper
and extend the analysis of the spin-precession dynamics to the general case.

The asymptotic forms are obtained for the initial state,
$Q\simeq Q_1^{\rm in}+Q_2^{\rm in}$
where
\begin{equation}
\label{TSSff-in}
Q_j^{\rm in}=
k_{j{\rm R}}{\rm sech}(\chi_{j{\rm R}}+\rho_j/2)\Pi_j
\e^{\smalli\chi_{j{\rm I}}}.
\end{equation}
and for the final state,
$Q\simeq Q_1^{\rm fin}+Q_2^{\rm fin}$
where
\begin{equation}
\label{TSSff-fin}
Q_j^{\rm fin}=
k_{j{\rm R}}{\rm sech}\left(\chi_{j{\rm R}}+\rho_j/2+s\right)
{\tilde \Pi}_j\e^{\smalli\chi_{j{\rm I}}}.
\end{equation}
Here we have defined
\begin{eqnarray}
s=\ln\left\{1-\frac{(k_1+k_1^*)(k_2+k_2^*)}{|k_1+k_2^*|^2}
\big|{\rm tr}\big(\Pi_1\Pi_2^\dagger\big)\big|\right\},
\end{eqnarray}
and, for $(j,l)=(1,2)$ or $(2,1)$,
\begin{eqnarray}
\label{new pi ff}
\tilde{\Pi}_j\eq
\e^{-s}\Bigg\{\Pi_j-\frac{k_l+k_l^*}{k_j+k_l^*}
\left(\Pi_j\Pi_l^\dagger\Pi_l+\Pi_l\Pi_l^\dagger\Pi_j\right)\nonumber\\
\espace\qquad{}+
\left(\frac{k_l+k_l^*}{k_j+k_l^*}\right)^2{\rm tr}\big(\Pi_j\Pi_l^\dagger
\big)\Pi_l
\Bigg\},
\end{eqnarray}
which are shown to be normalized in unity, $||\tilde{\Pi}_j||_2=1$.
Each polarization matrix $\Pi_j$ of a ferromagnetic soliton
can be expressed by three real variables $\tau_j,\theta_j,\varphi_j$ as
\begin{eqnarray}
\Pi_{j}=\e^{\smalli\tau_j}\left(\hspace*{-1mm}\begin{array}{cc}
\displaystyle
\cos^2\frac{\theta_j}{2}\e^{-\smalli\varphi_j} 
\hspace*{-1mm}&\hspace*{-1mm}
\displaystyle
\cos\frac{\theta_j}{2}\sin\frac{\theta_j}{2} \\
\displaystyle
\cos\frac{\theta_j}{2}\sin\frac{\theta_j}{2} 
\hspace*{-1mm}&\hspace*{-1mm}
\displaystyle
\sin^2\frac{\theta_j}{2}\e^{\smalli\varphi_j} 
\end{array}\hspace*{-1mm}\right).
\end{eqnarray}
In this expression,
the polarization matrices in the initial state $\Pi_j$
and in the final state ${\tilde \Pi}_j$
are given by
\begin{eqnarray}
\Pi_{j}\eq\e^{\smalli\tau_j}{\bf u}_j \cdot {\bf u}^T_j,\\
{\tilde\Pi}_{j}\eq
\e^{-s+\smalli\tau_j}{\tilde{\bf u}}_j\cdot{\tilde{\bf u}}^T_j,
\end{eqnarray}
where, with $(j, l)=(1,2)$, $(2,1)$,
\begin{eqnarray}
{\bf u}_{j}\eq\left(\begin{array}{cc}
\displaystyle
\cos{\frac{\theta_j}{2}}\e^{-\smalli\frac{\varphi_j}{2}} \\
\displaystyle
\sin{\frac{\theta_j}{2}}\e^{\smalli\frac{\varphi_j}{2}}
\end{array}\right),\\
\label{scat law}
{\tilde{\bf u}}_j\eq{\bf u}_j-\frac{k_l+k_l^*}{k_j+k_l^*}
\left({\bf u}^\dagger_l \cdot {\bf u}_j \right) {\bf u}_l.
\end{eqnarray}
This defines the collision law for the ferromagnetic-ferromagnetic soliton
collision.
Similar collision laws for the matrix KdV equation can be found
in ref. \citen{mtxKdV}.

We calculate the spins for each soliton to investigate their collision.
It will be interpreted as a spin precession around the total spin.
In the initial state, following eq.(\ref{OSspin}), we have the spin of
soliton-$j$ as
\begin{equation}
{\bf F}_{j}=2|k_{j{\rm R}}|\left(\begin{array}{ccc}
\sin{\theta_j}\cos{\varphi_j} \\
\sin{\theta_j}\sin{\varphi_j} \\
\cos{\theta_j} \end{array}\right).
\end{equation}
Thanks to the scattering law (\ref{scat law}),
the final state spins can be obtained through ${\bf F}_{1,2}$ by
\begin{eqnarray}
{\tilde{\bf F}}_{j}\eq\e^{-s}\Bigg[
\left\{1-\frac{2k_{l{\rm R}}(k_{1{\rm R}}+k_{2{\rm R}})}{|k_1+k_2^*|^2}
\right\}{\bf F}_{j}\nonumber\\
\espace\qquad{}+\frac{k_{2{\rm I}}-k_{1{\rm I}}}{|k_1+k_2^*|^2}
({\bf F}_{j}\times{\bf F}_{l})\nonumber\\
\label{finspin}
\espace\qquad{}+\left\{\e^{s}-1
+\frac{2k_{j{\rm R}}(k_{1{\rm R}}+k_{2{\rm R}})}{|k_1+k_2^*|^2}
\right\}{\bf F}_{l}\Bigg],
\end{eqnarray}
where
\begin{equation}
\e^{s}=1+\frac{|{\bf F}_{\rm T}|^2-(|{\bf F}_1|-|{\bf F}_2|)^2}
{4|k_{1}+k_{2}^*|^2}.
\end{equation}
The conserved total spin,
${\bf F}_{\rm T}\equiv{\bf F}_1+{\bf F}_2={\tilde{\bf F}}_1+{\tilde{\bf F}}_2$,
is given by
\begin{equation}
{\bf F}_{\rm T}=\left(\begin{array}{ccc}
2|k_{1{\rm R}}|\sin{\theta_1}\cos{\varphi_1}+
2|k_{2{\rm R}}|\sin{\theta_2}\cos{\varphi_2} \\
2|k_{1{\rm R}}|\sin{\theta_1}\sin{\varphi_1}+
2|k_{2{\rm R}}|\sin{\theta_2}\sin{\varphi_2} \\
2|k_{1{\rm R}}|\cos{\theta_1}+
2|k_{2{\rm R}}|\cos{\theta_2} \end{array} \right).
\end{equation}

Considering spin rotation around the total spin ${\bf F}_{\rm T}$, 
we can find `rotated spin' as
\begin{equation}
{\bf F}_{j}^{\rm rot}=f^{-1}(\varphi)h^{-1}(\theta)f(\omega)
h(\theta)f(\varphi){\bf F}_{j},
\end{equation}
where
\begin{eqnarray}
f(\varphi)\eq\left(\hspace*{-1mm}\begin{array}{ccc}
\cos{\varphi} \hspace*{-1mm}&\hspace*{-1mm} \sin{\varphi}
\hspace*{-1mm}&\hspace*{-1mm} 0 \\
-\sin{\varphi} \hspace*{-1mm}&\hspace*{-1mm}  \cos{\varphi}
\hspace*{-1mm}&\hspace*{-1mm} 0 \\
0 \hspace*{-1mm}&\hspace*{-1mm} 0 \hspace*{-1mm}&\hspace*{-1mm} 1
\end{array} \hspace*{-1mm}\right),\nonumber\\
h(\theta)\eq\left(\hspace*{-1mm}\begin{array}{ccc}
\cos{\theta} \hspace*{-1mm}&\hspace*{-1mm} 0
\hspace*{-1mm}&\hspace*{-1mm} -\sin{\theta} \\
0 \hspace*{-1mm}&\hspace*{-1mm} 1 \hspace*{-1mm}&\hspace*{-1mm} 0 \\
\sin{\theta} \hspace*{-1mm}&\hspace*{-1mm} 0
\hspace*{-1mm}&\hspace*{-1mm} \cos{\theta}
\end{array}\hspace*{-1mm} \right),
\end{eqnarray}
and
\begin{equation}
\cos{\varphi}=\frac{F^x_{\rm T}}{\sqrt{(F^x_{\rm T})^2
+(F^y_{\rm T})^2}},
\,\,
\sin{\varphi}=\frac{F^y_{\rm T}}{\sqrt{(F^x_{\rm T})^2
+(F^y_{\rm T})^2}},
\nonumber
\end{equation}
\begin{equation}
\cos{\theta}=\frac{F^z_{\rm T}}{|{\bf F}_{\rm T}|},
\,\,
\sin{\theta}=\frac{\sqrt{(F^x_{\rm T})^2+(F^y_{\rm T})^2}}
{|{\bf F}_{\rm T}|}.
\end{equation}
Then,
\begin{eqnarray}
{\bf F}_{j}^{\rm rot}\eq\frac{1}{2}\Bigg[
\left\{(1+\Delta_{jl})+(1-\Delta_{jl})\cos{\omega})\right\}{\bf F}_{j}\nonumber\\
\espace\qquad{}+
2\sin{\omega}\frac{({\bf F}_{j}\times{\bf F}_{l})}{|{\bf F}_{\rm T}|}
\nonumber\\
\label{rotspin}
\espace\qquad{}+
\left\{(1+\Delta_{jl})-(1+\Delta_{jl})\cos{\omega}\right\}{\bf F}_{l}\Bigg],
\end{eqnarray}
where
\begin{equation}
\Delta_{jl}=\frac{|{\bf F}_j|^2-|{\bf F}_l|^2}{|{\bf F}_{\rm T}|^2}.
\end{equation}
The rotation angle $\omega$ is determined by setting
${\bf F}_{1}^{\rm rot}={\tilde {\bf F}}_{1}$
through eqs.~(\ref{finspin}) and (\ref{rotspin}),
\begin{eqnarray}
\espace
\cos{\omega}=\frac{4(k_{1{\rm I}}-k_{2{\rm I}})^2-|{\bf F}_{\rm T}|^2}
{4(k_{1{\rm I}}-k_{2{\rm I}})^2+|{\bf F}_{\rm T}|^2},\\
\espace 
\sin{\omega}=\frac{4(k_{2{\rm I}}-k_{1{\rm I}})|{\bf F}_{\rm T}|}
{4(k_{1{\rm I}}-k_{2{\rm I}})^2+|{\bf F}_{\rm T}|^2}.
\end{eqnarray}

For the case that the amplitude and velocity of two ferromagnetic solitons
are same,
$|k_{\rm 1R}|=|k_{\rm 2R}|\equiv N_T/4$ and $|k_{\rm 1I}|=|k_{\rm 2I}|
=k_{\rm I}$, the final state spins $\tilde{\bf F}_j$ are given by
\begin{equation}
\tilde{\bf F}_j=\cos^2{\frac{\omega}{2}}{\bf F}_j+
\sin{\omega} 
\frac{({\bf F}_j \times {\bf F}_l)}{|{\bf F}_T|}+
\sin^2{\frac{\omega}{2}}{\bf F}_l,
\end{equation}
where $(j, l)=(1,2)$, $(2,1)$.
The rotation angle $\omega$ depends only on the ratio $k_I/k_R$
and the magnitude of the normalized total spin ${\mathcal F}\equiv
|{\bf F}_T/N_T|$ as
\begin{equation}
\label{acos}
\omega=2\arccos{\left(\left[1+\left(\frac{k_R}{k_I}\right)^2\, {\mathcal F}^2
\right]^{-1/2}\right)}.
\end{equation}
The principal value should be taken for the arccosine function:
$0\le \arccos{x}\le \pi$.

\begin{figure}[tbp]
  \begin{center}
    \includegraphics[height=.22\textheight]{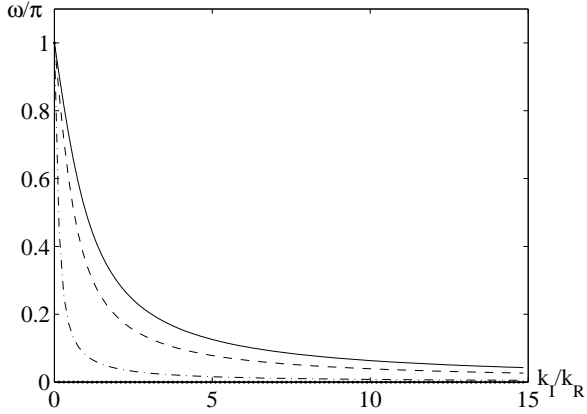}
  \end{center}
  \caption{Velocity dependence of the rotational angle
  in spin precession for the different initial relative
  angles between two spins, ${\mathcal F}=1$ (solid line),
  $0.5$ (dashed line), $0.0157\pi$ (dash-dot line)
  and $0$ (dotted line).}
  \label{fig:rot_angle}
\end{figure}

Setting $k_{\rm I}\gg k_{\rm R}$ in eq.~(\ref{acos}),
one gets the small rotation angle, $\omega\simeq 0$.
In the opposite case, $k_{\rm I}\ll k_{\rm R}$,
each spin of two colliding solitons almost reverses its orientation,
$\omega\simeq\pi$.
Recall that $k_{\rm I}$ is the speed of soliton. We can understand
these phenomena since a slower soliton spends the longer time inside
the collisional region.
Figure \ref{fig:rot_angle} shows the velocity dependence
of the rotation angle for various initial normalized spins.
When ${\mathcal F}=0$, which corresponds to the case
of antiparallel spin collision, the spin precession can not occur
as shown by the dotted line in Fig.\ref{fig:rot_angle}.

\begin{figure}[htbp]
  \begin{center}
   \includegraphics[height=.21\textheight]{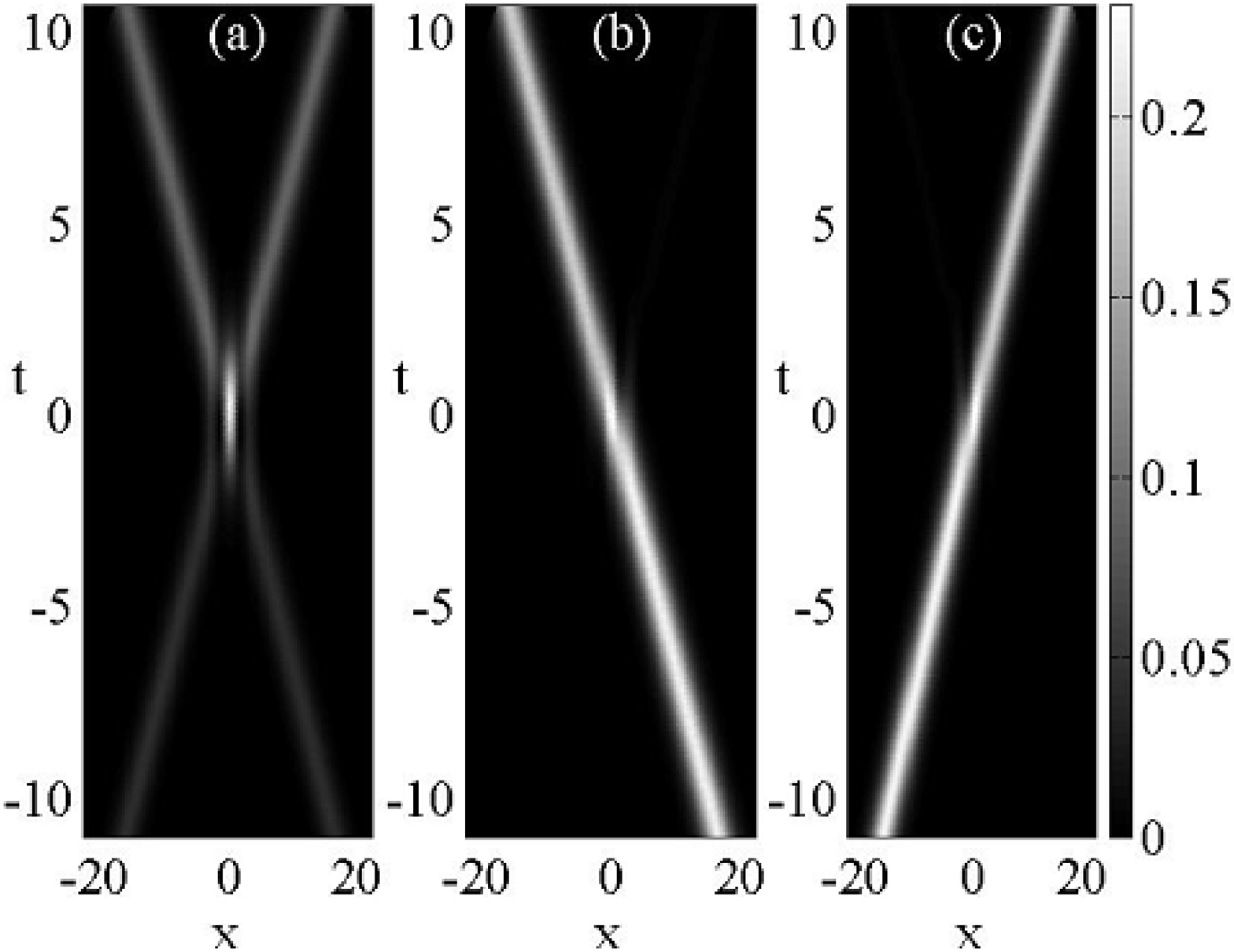}
  \end{center}
  \caption{Density plots of (a) $|\phi_0|^2$, (b) $|\phi_{1}|^2$ and (c)
  $|\phi_{-1}|^2$ for a fast ferromagnetic-ferromagnetic collision. 
  The parameters used here are $k_1=0.5-0.75\i$, $k_2=-0.5+0.75\i$,
  $\alpha_1=4/17$, $\beta_1=16/17$, $\gamma_1=1/17$,
  $\alpha_2=4/17$, $\beta_2=1/17$, $\gamma_2=16/17$.}
  \label{fig:ff_fast}
  \begin{center} 
   \includegraphics[height=.21\textheight]{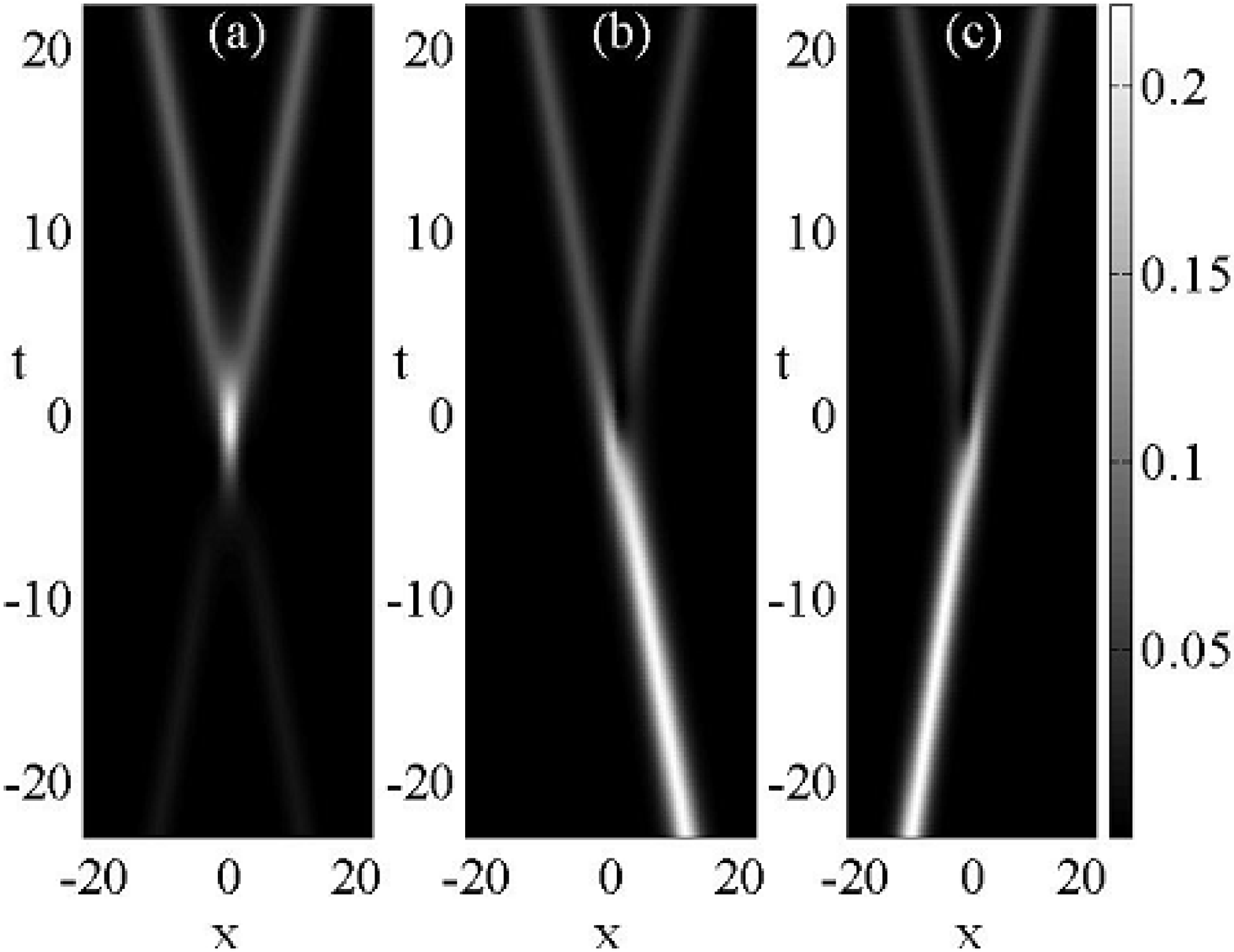}
  \end{center}
  \caption{Density plots of (a) $|\phi_0|^2$, (b) $|\phi_{1}|^2$ and (c)
  $|\phi_{-1}|^2$ for a medium speed ferromagnetic-ferromagnetic collision. 
  The parameters are the same as those of Fig.\ \ref{fig:ff_fast}
  except for $k_{1{\rm I}}=-0.25$, $k_{2{\rm I}}=0.25$.}
  \label{fig:ff_middle}
  \begin{center}
    \includegraphics[height=.21\textheight]{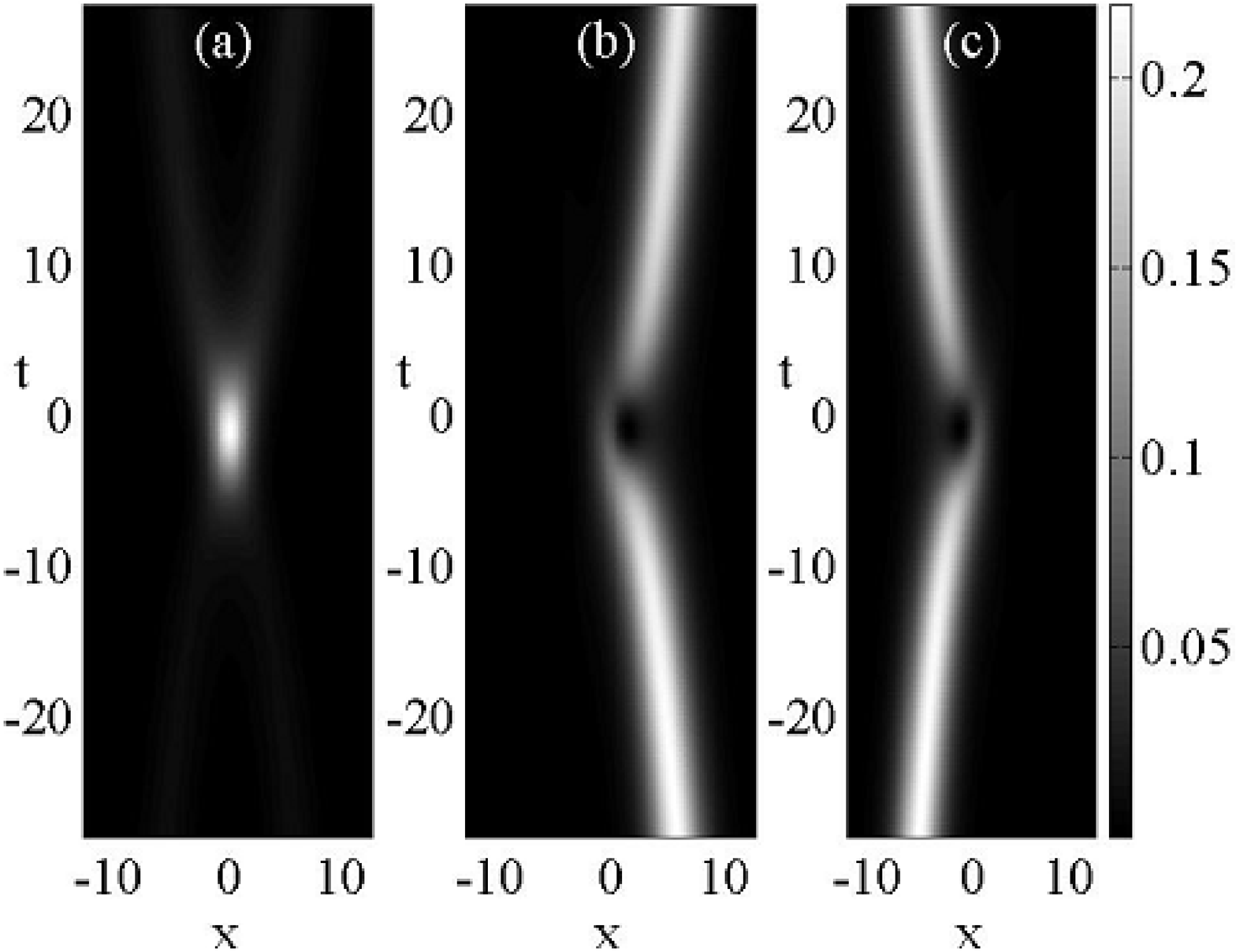}
  \end{center}
  \caption{Density plots of (a) $|\phi_0|^2$, (b) $|\phi_{1}|^2$ and (c)
  $|\phi_{-1}|^2$ for a slow ferromagnetic-ferromagnetic collision. 
  The parameters are the same as those of Fig.\ \ref{fig:ff_fast}
  except for $k_{1{\rm I}}=-0.05$, $k_{2{\rm I}}=0.05$.}
  \label{fig:ff_slow}
\end{figure}

In Fig.\ \ref{fig:ff_fast}--Fig.\ \ref{fig:ff_slow}, we give examples
of this type of collisions changing $k_{\rm I}$, with the other conditions
fixed, to illustrate the velocity dependence.
The initial normalized spin for the parameter set given
in the captions is ${\mathcal F}=0.5$.
The rotation angles are $\omega\simeq 0.2\pi$, $0.5\pi$ and $0.9\pi$ for
Fig.\ \ref{fig:ff_fast}, Fig.\ \ref{fig:ff_middle} and
Fig.\ \ref{fig:ff_slow}, respectively.
The internal shift $\phi_1 \to \phi_{-1}$, and vice versa, gradually
increase by slowing down the velocity of the solitons.

\section{\label{sec:conclusions}Conclusions and Discussions}
In the present paper, we have studied the soliton properties in
$F=1$ spinor Bose--Einstein condensates on the basis of
the integrable model introduced in ref. 23.

The complete classification of the one-soliton solution has been
carried out. There exist two distinct spin
states, ferromagnetic, $|{\mathbf F}_{\rm T}|=N_{\rm T}$ and
polar, $|{\mathbf F}_{\rm T}|=0$.
In the ferromagnetic state, the spatial part and the spinor part
of the soliton decouple (ferromagnetic soliton).
In the polar state, dissimilar shaped solitons which we call
polar soliton for ${\mathbf f}(x)=0$ and split soliton otherwise
are energetically degenerate.
The polar soliton has one peak and the space--spinor decoupling holds.
On the other hand, a split soliton consists of twin peaks and the
profiles of the three components are different.
Changing the polarization parameters, one may control
the distance between these peaks continuously.

These properties remind us of the Sasa--Satsuma higher-order nonlinear
Schr\"odinger (HNLS) equation, which is a unique soliton equation
generating solitons with two peaks.
In fact, the spinor model and the Sasa--Satsuma model, though the former
is three-component and the latter is one-component, have exactly the same
form of the envelope, i.e., the density (\ref{polar-num}).
(See also eq.(52a) in ref. \citen{SasaSatsuma}.)
The relation between the multi-component NLS equation and
the one-component HNLS equation is interesting and
remains as a future problem.

We have also shown explicit two-soliton solutions which rule
collisional phenomena of the multiple solitons.
Specifying the initial conditions, we have demonstrated
two-soliton collisions in three characteristic cases:
polar-polar, polar-ferromagnetic, ferromagnetic-ferromagnetic.
In their collisions, the polar soliton is always ``passive" which means
that it can not rotate its partner's polarization while
the ferromagnetic soliton does.
Thus, in the polar-ferromagnetic collision, one can use the polar
soliton as a signal and ferromagnetic soliton as a switch
to realize a coherent matter-wave switching device.
Collision of two ferromagnetic solitons can be interpreted as the
spin precession around the total spin. The rotation angle
depends on the total spin, amplitude and velocity of the solitons.
Only varying the velocity induces drastic change of
the population shifts among the components.

Remark that in current experimental setups~\cite{spindynamics_ex},
the initial population among the three components can be controlled
by magnetic field gradients during preparing condensates.
The components overlapping in solitons can be spatially
separated by applying a weak Stern--Gerlach field before a time-of-flight
measurement.
We hope that those phenomena predicted in this work are observed in
experiments and open up a variety of applications in coherent atom transport,
and quantum information.

\section*{Acknowledgment}
One of the authors (J.\ I.) would like to thank T.~Tsuchida
for many useful discussions.

\onecolumn
\appendix
\renewcommand{\theequation}
{\Alph{section}$\cdot$\arabic{equation}}
\section{Two-Soliton Solution of Spinor Model}
\label{sec:2-soliton}
\setcounter{equation}{0}
In this appendix, we write out general two-soliton solution given by
(\ref{N soliton}) with $N=2$. According to (\ref{S matrix}),
the matrix $S$ in this case takes the form:
\begin{equation}
\label{S matrix 2}
\displaystyle
S=\left[
\begin{array}{rr}
\displaystyle
I+\frac{M_{11}}{k_1+k_1^*}\e^{\chi_1+\chi_1^*}
+\frac{M_{12}}{k_1+k_2^*}\e^{\chi_1+\chi_2^*} &
\displaystyle
\frac{M_{11}}{k_2+k_1^*}\e^{\chi_1+\chi_1^*}
+\frac{M_{12}}{k_2+k_2^*}\e^{\chi_1+\chi_2^*}\\
\displaystyle
\frac{M_{21}}{k_1+k_1^*}\e^{\chi_2+\chi_1^*}
+\frac{M_{22}}{k_1+k_2^*}\e^{\chi_2+\chi_2^*} &
\displaystyle
I+\frac{M_{21}}{k_2+k_1^*}\e^{\chi_2+\chi_1^*}
+\frac{M_{22}}{k_2+k_2^*}\e^{\chi_2+\chi_2^*}
\end{array}
\right],
\end{equation}
where $2\times 2$ matrices $M_{ij}$ are
\begin{equation}
M_{ij}=
\displaystyle
\left(
\begin{array}{cc}
\kappa_{ij} & \lambda_{ij} \\
\mu_{ij} & \nu_{ij}
\end{array}
\right),
\end{equation}
with
\begin{eqnarray}
\kappa_{ij} = \frac{\alpha_i\alpha_j^*+\beta_i\beta_j^*}{k_i+k_j^*},\quad
\lambda_{ij}= \frac{\alpha_i\gamma_j^*+\beta_i\alpha_j^*}{k_i+k_j^*},\quad
\mu_{ij}    = \frac{\gamma_i\alpha_j^*+\alpha_i\beta_j^*}{k_i+k_j^*},\quad
\nu_{ij}    = \frac{\alpha_i\alpha_j^*+\gamma_i\gamma_j^*}{k_i+k_j^*}.
\end{eqnarray}
Here the elements of the polarization matrices $\alpha_j$, $\beta_j$, $\gamma_j$,
and the spectral parameters $k_j$ are determined by the initial state
($t\to-\infty$). Recall that all $x$ and $t$ dependence is only through the
complex coordinates $\chi_j=\chi_j(x,t)$.

The determinant of (\ref{S matrix 2}) is
\begin{eqnarray}
\label{det2}
\det S\eq
1+\sum_{j,l}A_{jl}\e^{\chi_j+\chi_l^*}+
\sum_{i,j,k,l}B_{ijkl}\e^{\chi_i+\chi_j^*+\chi_k+\chi_l^*} \nonumber\\
\espace+
\sum_{i,j,k,l}C_{ijkl}\e^{2\chi_i+2\chi_j^*+\chi_k+\chi_l^*}+
D\e^{2(\chi_1+\chi_1^*+\chi_2+\chi_2^*)},
\end{eqnarray}
where
\begin{eqnarray}
A_{ij}\eq \frac{{\rm tr}M_{jl}}{k_j+k_l^*},\\
B_{ijkl}\eq \frac{\det\left(
\begin{array}{cc}
\kappa_{ij} & \lambda_{ij} \\
\mu_{kl} & \nu_{kl}
\end{array}
\right)+\frac{1}{2}\left\{{\rm tr}(M_{ij}M_{kl})-{\rm tr}(M_{il}M_{kj})
 \right\}}{(k_i+k_j^*)(k_k+k_l^*)},\\
C_{ijkl}\eq \frac{1}{k_i+k_j^*}\left\{
\frac{1}{(k_i+k_j^*)(k_k+k_l^*)}-\frac{1}{(k_i+k_l^*)(k_k+k_j^*)}
\right\} \nonumber\\
\espace \mbox{}\times \left\{ \det\left(
\begin{array}{cc}
M_{ij} & \begin{array}{c} \lambda_{il} \\ \nu_{il} \end{array} \\
\begin{array}{cc} \mu_{kj} & \nu_{kj} \end{array} & \nu_{kl}
\end{array}\right) + \det\left( \begin{array}{cc}
\kappa_{kl} & \begin{array}{cc} \kappa_{kj} & \lambda_{kj} \end{array} \\
\begin{array}{c} \kappa_{il} \\ \mu_{il} \end{array} & M_{ij}
\end{array}\right)
\right\},\\
D\eq \left\{
\frac{|k_1-k_2|^2}{(k_1+k_1^*)(k_2+k_2^*)|k_1+k_2^*|^2}
\right\}^2\det\left(\begin{array}{cc}
M_{11} &M_{12} \\
M_{21} &M_{22} 
\end{array}
\right).
\end{eqnarray}
Then, the two-soliton solution is given by
\begin{eqnarray}
\label{2SS}
Q\eq\frac{1}{\det S}\left\{ (\tilde{S}_{11}+\tilde{S}_{21})\Pi_1\e^{\chi_1}+
  (\tilde{S}_{12}+\tilde{S}_{22})\Pi_2\e^{\chi_2}\right\}\nonumber \\
  \eq\frac{1}{\det S}\left\{ \left(
  \begin{array}{cc}(\tilde{S}_{11}^{11}+\tilde{S}_{21}^{11})\beta_1+
  (\tilde{S}_{11}^{12}+\tilde{S}_{21}^{12})\alpha_1 &
  (\tilde{S}_{11}^{11}+\tilde{S}_{21}^{11})\alpha_1+
  (\tilde{S}_{11}^{12}+\tilde{S}_{21}^{12})\gamma_1 \\
  (\tilde{S}_{11}^{21}+\tilde{S}_{21}^{21})\beta_1+
  (\tilde{S}_{11}^{22}+\tilde{S}_{21}^{22})\alpha_1 &
  (\tilde{S}_{11}^{21}+\tilde{S}_{21}^{21})\alpha_1+
  (\tilde{S}_{11}^{22}+\tilde{S}_{21}^{22})\gamma_1
  \end{array}\right)\e^{\chi_1} \right. \nonumber \\
  \espace \qquad\qquad\mbox{}+\left.\left(
  \begin{array}{cc}(\tilde{S}_{12}^{11}+\tilde{S}_{22}^{11})\beta_2+
  (\tilde{S}_{12}^{12}+\tilde{S}_{22}^{12})\alpha_2 &
  (\tilde{S}_{12}^{11}+\tilde{S}_{22}^{11})\alpha_2+
  (\tilde{S}_{12}^{12}+\tilde{S}_{22}^{12})\gamma_2 \\
  (\tilde{S}_{12}^{21}+\tilde{S}_{22}^{21})\beta_2+
  (\tilde{S}_{12}^{22}+\tilde{S}_{22}^{22})\alpha_2 &
  (\tilde{S}_{12}^{21}+\tilde{S}_{22}^{21})\alpha_2+
  (\tilde{S}_{12}^{22}+\tilde{S}_{22}^{22})\gamma_2
  \end{array}\right)\e^{\chi_2}\right\}.
\end{eqnarray}
Here we use the tilde to denote cofactors. The cofactor $\tilde{S}_{ij}^{kl}$
is obtained by deleting the $(2i+k-2)$-th row and the $(2j+l-2)$-th column
in the determinant of $S$ and multiplying it by $(-1)^{k+l}$.
For instance, the cofactors $\tilde{S}_{1j}^{1l}$ read
\begin{eqnarray}
\tilde{S}_{11}^{11} \eq
1+ \sum_{j=1,2}\left\{\frac{\nu _{1j}}{k_1+k_j^*} \e^{\chi_1+\chi_j^*} 
 + \frac{{\rm tr}M_{2j}}{k_2+k_j^*}\e ^{\chi_2+\chi_j^*}
 +\frac{\det M_{2j}}{(k_2+k_j^*)^2}\e^{2\chi_2+2\chi_j^*}\right\}\nonumber \\
\espace \mbox{}+\sum_{j=1,2}\frac{\kappa_{2j} \nu_{1j} - \lambda_{2j} \mu_{1j} }
{(k_1+k_j^*)(k_2+k_j^*)} \e ^{\chi_1+\chi_2+2\chi_j^*}
+\frac{\kappa_{21} \nu_{22} - \lambda_{21} \mu_{22} +
\kappa_{22} \nu_{21} - \lambda_{22} \mu_{21} }
{(k_2+k_1^*)(k_2+k_2^*)}\e ^{2\chi_2+\chi_1^*+\chi_2^*}
\nonumber\\
\espace\mbox{}+ \left\{\frac{\kappa_{22} \nu_{11} - \lambda_{21} \mu_{12}+
\nu_{11} \nu_{22} - \nu_{12} \nu_{21} }{(k_1+k_1^*)(k_2+k_2^*)}
+ \frac{\kappa_{21} \nu_{12} - \lambda_{22} \mu_{11}+
\nu_{12} \nu_{21} - \nu_{11} \nu_{22} }{(k_1+k_2^*)(k_2+k_1^*)}\right\}
\e ^{\chi_1+\chi_2+\chi_1^*+\chi_2^*}    \nonumber\\
\espace\mbox{}+ \frac{|k_1-k_2|^2}{(k_1+k_1^*)(k_2+k_2^*)|k_1+k_2^*|^2}
         \sum_{j=1,2}
         \frac{1}{k_2+k_j^*}
         \det\left(
         \begin{array}{ccc}
         \nu _{11} & \mu _{1j} & \nu _{12}\\
         \lambda _{21} & \kappa_{2j} &\lambda_{22}\\
         \nu _{21} & \mu _{2j} & \nu _{22}
         \end{array}
         \right)
         \e ^{\chi_1+2\chi_2+\chi_1^*+\chi_2^*+\chi_j^*},\\
\tilde{S}_{11}^{12} \eq
\sum_{j=1,2}\left\{-\frac{\lambda _{1j}}{k_1+k_j^*} \e^{\chi_1+\chi_j^*} 
+\frac{\kappa_{1j} \lambda_{2j} - \kappa_{2j} \lambda_{1j} }
{(k_1+k_j^*)(k_2+k_j^*)} \e ^{\chi_1+\chi_2+2\chi_j^*}\right\}
\nonumber\\
\espace\mbox{}+\left\{\frac{\kappa_{12} \lambda_{21} - \kappa_{22} \lambda_{11} +
\lambda_{12} \nu_{21} -\lambda_{11} \nu_{22} }{(k_1+k_1^*)(k_2+k_2^*)}
+ \frac{\kappa_{11} \lambda_{22} - \kappa_{21} \lambda_{12} +
\lambda_{11} \nu_{22} - \lambda_{12} \nu_{21} }{(k_1+k_2^*)(k_2+k_1^*)}\right\}
\e ^{\chi_1+\chi_2+\chi_1^*+\chi_2^*}    \nonumber\\
\espace\mbox{}- \frac{|k_1-k_2|^2}{(k_1+k_1^*)(k_2+k_2^*)|k_1+k_2^*|^2}
         \sum_{j=1,2}
         \frac{1}{k_2+k_j^*}
         \det\left(
         \begin{array}{ccc}
         \lambda_{11} & \kappa_{1j} & \lambda_{12}\\
         \lambda _{21} & \kappa_{2j} &\lambda_{22}\\
         \nu _{21} & \mu _{2j} & \nu _{22}
         \end{array}
         \right)
         \e ^{\chi_1+2\chi_2+\chi_1^*+\chi_2^*+\chi_j^*},\\
\tilde{S}_{12}^{11} \eq
-\sum_{j=1,2}\left\{\frac{\kappa_{1j}}{k_2+k_j^*} \e^{\chi_1+\chi_j^*} 
+\frac{\det M_{1j}}{(k_1+k_j^*)(k_2+k_j^*)} \e ^{2\chi_1+2\chi_j^*}\right\}
\nonumber\\
\espace\mbox{}
-\sum_{j=1,2}\frac{\kappa_{1j} \nu_{2j} - \lambda_{1j} \mu_{2j} }
{(k_2+k_j^*)(k_2+k_j^*)} \e ^{\chi_1+\chi_2+2\chi_j^*}
- \frac{\kappa_{11} \nu_{22} - \lambda_{11} \mu_{22} +
\kappa_{12} \nu_{21} - \lambda_{12} \mu_{21} }
{(k_2+k_1^*)(k_2+k_2^*)}\e ^{\chi_1+\chi_2+\chi_1^*+\chi_2^*}
\nonumber\\
\espace\mbox{}
- \left\{\frac{\kappa_{12}\nu_{11} - \lambda_{11}\mu_{12}}
{(k_1+k_1^*)(k_2+k_2^*)}
+ \frac{\kappa_{11} \nu_{12} - \lambda_{12} \mu_{11}}
{(k_1+k_2^*)(k_2+k_1^*)}\right\}
\e ^{2\chi_1+\chi_1^*+\chi_2^*}
\nonumber\\
\espace\mbox{}- \frac{|k_1-k_2|^2}{(k_1+k_1^*)(k_2+k_2^*)|k_1+k_2^*|^2}
         \sum_{j=1,2}
         \frac{1}{k_2+k_j^*}
         \det\left(
         \begin{array}{ccc}
         \kappa_{1j} & \lambda_{11} & \lambda_{12}\\
         \mu _{1j} & \nu _{11} & \nu _{12}\\
         \mu _{2j} & \nu _{21} & \nu _{22}
         \end{array}
         \right)
         \e ^{2\chi_1+\chi_2+\chi_1^*+\chi_2^*+\chi_j^*},\\
\tilde{S}_{12}^{12} \eq
\sum_{j=1,2}\left\{-\frac{\lambda _{1j}}{k_2+k_j^*} \e^{\chi_1+\chi_j^*} 
+\frac{ \kappa_{1j} \lambda_{2j} - \kappa_{2j} \lambda_{1j}}
{(k_2+k_j^*)(k_2+k_j^*)} \e ^{\chi_1+\chi_2+2\chi_j^*}\right\}
\nonumber\\
\espace\mbox{}+ \left\{\frac{\lambda_{11} \nu_{12} - \lambda_{12} \nu_{11}}
{(k_1+k_1^*)(k_2+k_2^*)}
+ \frac{\lambda_{12} \nu_{11} - \lambda_{11} \nu_{12}}
{(k_1+k_2^*)(k_2+k_1^*)}\right\}
\e ^{2\chi_1+\chi_1^*+\chi_2^*}    \nonumber\\
\espace\mbox{}+ \frac{\kappa_{11} \lambda_{22} - \kappa_{22} \lambda_{11} +
\kappa_{12} \lambda_{21} - \kappa_{21} \lambda_{12} }
{(k_2+k_1^*)(k_2+k_2^*)}\e ^{\chi_1+\chi_2+\chi_1^*+\chi_2^*}
\nonumber\\
\espace\mbox{}+ \frac{|k_1-k_2|^2}{(k_1+k_1^*)(k_2+k_2^*)|k_1+k_2^*|^2}
         \sum_{j=1,2}
         \frac{1}{k_2+k_j^*}
         \det\left(
         \begin{array}{ccc}
         \kappa_{1j} & \lambda_{11} & \lambda_{12}\\
         \mu _{1j} & \nu _{11} & \nu _{12}\\
         \kappa_{2j} & \lambda _{21} & \lambda _{22}
         \end{array}
         \right)
         \e ^{2\chi_1+\chi_2+\chi_1^*+\chi_2^*+\chi_j^*}.
\end{eqnarray}

The determinant of matrix $S$ (\ref{det2}) is a polynomial in $\e^{\chi_j}$
and $\e^{\chi_j^*}$ of degree 8, and the cofactors are degree 6,
giving rise to a variety of the scattering processes.
This property contrasts to the two-soliton solution of the general
$m$-component Manakov model, in which the determinant of $S$
and the cofactors are degree 4 and 2, respectively. 
If one set $\det\Pi_j=0\,\,(j=1,2)$, corresponding to the ferromagnetic state,
the determinants in the two higher degree's coefficients
vanish, and the expansions of $\det S$ and $\tilde{S}_{ij}^{kl}$ terminate at
degree 4 and 2, respectively.

\vspace*{-5mm}
\begin{thebibliography}{99}
\bibitem{refsoliton}
M. J. Ablowitz and H. Segur:
\emph{Solitons and the Inverse Scattering Transform}
(SIAM, Philadelphia, 1981).

\bibitem{SolRice}
Kevin E. Strecker, Guthrie B. Partridge, Andrew G. Truscott,
Randall G. Hulet: Nature \/(London) \textbf{417} (2002) 150.

\bibitem{SolEns}
L. Khaykovich, F. Schreck, G. Ferrari, T. Bourdel, J. Cubizolles,
L. D. Carr, Y. Castin and C. Salomon: Science \textbf{296} (2002) 1290.

\bibitem{darksolex}
S. Burger, K. Bongs, S. Dettmer, W. Ertmer and K. Sengstock:
Phys. Rev. Lett. \textbf{83} (1999) 5198,
J. Denschlag, \emph{et al.}: Science \textbf{287} (2000) 97.

\bibitem{Khawaja}
U. A. Khawaja, H. T. C. Stoof, R. G. Hulet, K. E. Strecker and G. B. Partridge:
Phys. Rev. Lett. \textbf{89} (2002) 200404.

\bibitem{Salasnich}
L. Salasnich, A. Parola and L. Reatto: Phys. Rev. \/A \textbf{66} (2002) 043603;
Phys. Rev. Lett. \textbf{91} (2003) 080405.

\bibitem{Leung}
V. Y. F. Leung, A. G. Truscott and K. G. H. Baldwin:
Phys. Rev. \/A \textbf{66} (2002) 061602.

\bibitem{Carr}
L. D. Carr and Y. Castin: Phys. Rev. \/A \textbf{66} (2002) 063602,
L. D. Carr and J. Brand: Phys. Rev. Lett. \textbf{92} (2004) 040401.

\bibitem{binaryBEC}
C. J. Myatt, E. A. Burt, R. W. Ghrist, E. A. Cornell and C. E. Wieman CE:
Phys. Rev. Lett. \textbf{78} (1997) 586.

\bibitem{spinor_BEC_MIT}
D. M. Stamper-Kurn, M. R. Andrews, A. P. Chikkatur, S. Inouye,
H.-J. Miesner, J. Stenger and W. Ketterle:
Phys. Rev. Lett. \textbf{80} (1998) 2027,
H.-J. Miesner, D. M. Stamper-Kurn, J. Stenger, S. Inouye,
A. P. Chikkatur and W. Ketterle: Phys. Rev. Lett. \textbf{82} (1999) 2228.

\bibitem{Meystre}
Pierre Meystre: \emph{Atom Optics},
(Springer-Verlag, New York, Inc., 2001).

\bibitem{2component}
For example, see P. G. Kevrekidis, H. E. Nistazakis, D. J. Frantzeskakis,
B. A. Malomed and R. Carretero-Gonzaalez R: Eur. Phys. J. D \textbf{28}
(2004) 181,
D. Schumayer and B. Apagyi: Phys. Rev. \/A \textbf{69} (2004) 043620
and references therein.

\bibitem{Soljacic}
M. Solja\v{c}i\'{c}, K. Steiglitz, S. M. Sears, M. Segev, M. H. Jakubowski
and R. Squier: Phys. Rev. Lett. \textbf{90} (2003) 254102.

\bibitem{opt_soliton_review}
A. V. Buryak, P. D. Trapani, D. V. Skryabin and S. Trillo:
Phys. Rep. \textbf{370} (2002) 63.

\bibitem{Ho}
Tin-Lun Ho: Phys. Rev. Lett. \textbf{81} (1998) 742.

\bibitem{Ohmi}
T. Ohmi and K. Machida: J. Phys. Soc. Jpn. \textbf{67} (1998) 1822.

\bibitem{Law}
C. K. Law, H. Pu and N. P. Bigelow: Phys. Rev. Lett. \textbf{81} (1999) 5257.

\bibitem{Koashi}
M. Koashi and M. Ueda: Phys. Rev. Lett. \textbf{84} (2000) 1066.

\bibitem{Ciobanu}
C. V. Ciobanu, S.-K. Yip and Tin-Lun Ho:
Phys. Rev. \/A \textbf{61} (2000) 033607.

\bibitem{Ueda}
M. Ueda and M. Koashi: Phys. Rev. \/A \textbf{65} (2000) 063602.

\bibitem{spindynamics_th}
H. Pu, C. K. Law, S. Raghavan, J. H. Eberly and N. P. Bigelow:
Phys. Rev. A \textbf{60} (1999) 1463.

\bibitem{spindynamics_ex}
H. Schmaljohann, M. Erhard, J. Kronjager, M. Kottke, S. van Staa,
L. Cacciapuoti, J. J. Arlt, K. Bongs and K. Sengstock:
Phys. Rev. Lett. \textbf{92} (2004) 040402;
M.-S. Chang, C. D. Hamley, M. D. Barrett, J. A. Sauer, K. M. Fortier,
W. Zhang, L. You and M. S. Chapman:
Phys. Rev. Lett. \textbf{92} (2004) 140403.

\bibitem{IMWlett}
J. Ieda, T. Miyakawa and M. Wadati: e-print cond-mat/0404569,
to appear in Phys. Rev. Lett. \textbf{93} (2004).

\bibitem{Olshanii}
M. Olshanii: Phys. Rev. Lett. \textbf{81} (1998) 938.

\bibitem{Mag1}
S. Inoue, M. R. Andrews, J. Stenger, H.-J. Miesner, D. M. Stamper-Kurn
and W. Ketterle: \textit{Nature} \/(London) \textbf{392} (1998) 151.

\bibitem{Mag2}
S. L. Cornish, N. R. Claussen, J. L. Roberts, E. A. Cornell
and C. E. Wieman: Phys. Rev. Lett. \textbf{85} (2000) 1795.

\bibitem{Opt1}
E. K. Fatemi, E. M. Jonse and P. D. Lett:
Phys. Rev. Lett. \textbf{85} (2000) 4462,

\bibitem{Opt2}
Jordan M. Gerton, Brian J. Frew and Randall G. Hulet:
Phys. Rev. \/A \textbf{64} (2001) 053410.

\bibitem{Opt3}
M. Theis, G. Thalhammer, K. Winkler, M. Hellwig, G. Ruff, R. Grimm,
J. H. Denschlag
: cond-mat/0404514.

\bibitem{Tsuchida1}
T. Tsuchida and M. Wadati: J. Phys. Soc. Jpn. \textbf{67} (1998) 1175,
and references therein.

\bibitem{Manakov}
S. V. Manakov: Sov. Phys.--JETP \textbf{38} (1974) 248.
For further extensions, see M.\ Hisakado and M.\ Wadati:
J. Phys. Soc. Jpn. \textbf{64} (1995) 408.

\bibitem{Rad}
R. Radhakrishnan, M. Lakshmanan and J. Hietarinta: 
Phys. Rev. \/E \textbf{56} (1997) 2213.

\bibitem{Tsuchida}
T. Tsuchida: Prog. Theor. Phys. \textbf{111} (2004) 151.

\bibitem{Saha}
R. Sahadevan, K. M. Tamizhmani and M.\ Lakshmanan:
J. Phys. \/A \textbf{19} (1986) 1783.

\bibitem{mtxKdV}
V. M. Goncharenko: Theor. Math. Phys. \textbf{126} (2001) 81.

\bibitem{SasaSatsuma}
N. Sasa and J. Satsuma: J. Phys. Soc. Jpn. \textbf{60} (1991) 409.

\end{thebibliography}
\end{document}